\documentclass[]{aastex631}
\usepackage{CJK}
\usepackage{amsmath}
\usepackage{natbib}
\graphicspath{{./}{figures/}}
\begin{document}
\begin{CJK*}{UTF8}{gbsn}

\title{Stellar Chromospheric Activity Database of Solar-like Stars Based on the LAMOST Low-Resolution Spectroscopic Survey III. Calibrating the Chromospheric Basal Flux and the Connection to Stellar Rotation}

\correspondingauthor{Weitao Zhang}
\email{zhangwt@zuwe.edu.cn}

\author{Weitao Zhang}
\affiliation{Nanxun Innovation Institute, Zhejiang University of Water Resources and Electric Power, Hangzhou, 310018, China}

\author{Han He}
\affiliation{National Astronomical Observatories, Chinese Academy of Sciences, Beijing 100101, China}
\affiliation{University of Chinese Academy of Sciences, Beijing 100049, China}

\author{Jun Zhang}
\affiliation{School of Physics and Optoelectronics Engineering, Anhui University, Hefei 230601, China}

\begin{abstract}
Based on the Ca\,II H and K lines observed by LAMOST, we employ the photospheric ($R'_{\rm HK}$) and basal ($R^+_{\rm HK,L}$) flux calibrated chromospheric activity indices to examine the relationship between chromospheric activity and the stellar rotation rate.
We identify the rotation periods of 11,108 stars observed by Kepler and TESS by cross-matching our chromospheric activity catalog with previous studies.
Our statistical results show that chromospheric activity increases with the rotation rate until it reaches a saturation level.
As the stellar effective temperature increases from 4950 to 5850 K, the saturation values of the rotation period ($P_{\rm rot}$) vary correspondingly from 4.38 to 1.23 days for $R'_{\rm HK}$ and from 9.88 to 1.33 days for $R^+_{\rm HK,L}$. Similarly, the corresponding saturation Rossby number Ro ranges from 0.200 to 0.032 for $R'_{\rm HK}$ and from 0.302 to 0.107 for $R^+_{\rm HK,L}$.
The saturation is also found to be significant in stars with thick convective zones, whereas it is less apparent in stars with higher effective temperatures.
For solar-like stars in the $T_{\rm eff}$ range of 4800 to 6000 K, The values of chromospheric activity indicators are saturated when $P_{\rm rot}<1.45 $ days (Ro$<$0.100) and $P_{\rm rot}<2.85 $ days (Ro$<$0.097) for $R'_{\rm HK}$ and $R^+_{\rm HK,L}$, respectively.

\end{abstract}


\section{Introduction}\label{sec:introduction}

Stars with convective outer layers universally exhibit magnetic activity.
Stellar magnetic activity is a fundamental manifestation of the dynamo process, influencing stellar evolution and structure.
Stellar activity levels are generally anti-correlated with both rotation period and age \citep{2022A&A...662A..41R}.
Solar-like and low-mass stars exhibit a clear correlation between their rotation rates and non-thermal emission levels, as measured in indicators including X-rays \citep{2003A&A...397..147P, 2011ApJ...743...48W, 2014ApJ...794..144R, 2018MNRAS.479.2351W} , ${\rm H}\alpha$ \citep{2014ApJ...795..161D,2017ApJ...834...85N,2023ApJS..264...12H, 2025ApJS..281....5Z} and Ca\,\textsc{ii} H and K  lines \citep{1984ApJ...279..763N,2017A&A...600A..13A, 2018A&A...618A..48M, 2022ApJ...929...80B}. In addition, stellar age can be estimated from stellar rotation and chromospheric activity. \citep{2008ApJ...687.1264M}. 

As the rotation rate increases, the magnetic activity in the stellar chromosphere does not increase monotonically.
For most rapidly rotating stars, there is a saturated regime of magnetic activity that appears related to Rossby number (Ro).
The dimensionless number Ro is the ratio of the stellar rotation period ($P_{\rm rot}$) to the convective turnover time ($\tau_c$). 
The saturation of stellar magnetic activity is relatively clearly seen in X-rays \citep{2003A&A...397..147P, 2011ApJ...743...48W, 2014ApJ...794..144R, 2018MNRAS.479.2351W}. Saturation is typically observed at Ro value of approximately 0.1 \citep{2003A&A...397..147P}, and generally less than 0.13 \citep{2011ApJ...743...48W}.
\citet{2008ApJ...687.1264M} investigated the activity-rotation relation for F7-K2 dwarfs, and suggested that the saturated X-ray appear for Ro$<$0.5, which corresponding to $P_{\rm rot}<6$ days for G2 dwarf.
The saturated phenomenon is widely studied in M-type stars, which possess thicker convective zones \citep{2003A&A...397..147P, 2014ApJ...795..161D,2017ApJ...834...85N, 2017A&A...600A..13A, 2018MNRAS.479.2351W, 2022ApJ...929...80B}.
In contrast, saturation is absent in late F-type stars \citep{2011ApJ...743...48W}.
some research reveals that a common dynamo mechanism operates in both giants and dwarfs \citep{1987A&A...177..131R, 2020NatAs...4..658L, 2024ApJ...977..138H}.
The saturation phenomenon exhibits a complex dependence on stellar age, mass, and other intrinsic stellar properties.

The Ca\,\textsc{ii} H and K lines are generally employed to indicate stellar magnetic activity. 
The chromospheric activity indicator known as the $S$-index, constructed from relative Ca\,\textsc{ii} H and K line core emissions at Mount Wilson Observatory (MWO), enables long-term monitoring of stellar activity cycles. \citep{1968ApJ...153..221W, 1978PASP...90..267V}.
The $S$-index, defined as the emission flux relative to the pseudo-continuum flux (a temperature-dependent quantity), has limitations when comparing stars of different spectral types. 
The stellar surface flux in the Ca\,\textsc{ii} H and K line cores can be derived from the $S$-index and is subsequently normalized by the bolometric flux to yield the $R_{\rm HK}$ index \citep{1982A&A...107...31M, 1984A&A...130..353R}.
The $R'_{\rm HK}$, defined by subtracting the photospheric contribution from $R_{\rm HK}$, is derived and has become the widely adopted indicator of chromospheric activity.
The Ca\,\textsc{ii} H and K lines naturally exhibit a basal chromospheric flux, which corresponds to the chromospheric heating in entirely inactive stars and demonstrates significant sensitivity to effective temperature \citep{1987A&A...172..111S, 2014MNRAS.445..270P, 2013A&A...549A.117M}.
A weaker dependence of the chromospheric activity indicator on luminosity class is preferable.
\citep{2013A&A...549A.117M} based on the lower envelope of $R'_{\rm HK}$ empirically construct the basal chromospheric flux calibrated index $R^+_{\rm HK}$.
In this work, we employ the classical chromospheric activity indicator $R'_{\rm HK}$ to investigate the activity-rotation relation, with parallel implementation of the $R^+_{\rm HK}$ index to evaluate potential improvements.

The Large Sky Area Multi-Object Fiber Spectroscopic Telescope (LAMOST, also called the Guoshoujing Telescope) has released massive spectral data since its pilot survey started in 2011 \citep{2012RAA....12.1197C, 2012RAA....12..723Z, 2012RAA....12.1243L}.
Extensive wavelength coverage and large-scale data are highly advantageous for statistical research on stellar chromosphere activity \citep{2022ApJS..263...12Z, 2023Ap&SS.368...63H, 2023ApJS..264...12H, 2024ApJS..271...19Y, 2024A&A...688A..23Z}.
We derived out the phtospheric and chromospheric basal flux from the Ca\,\textsc{ii} H and K lines observed in LAMOST The Low-Resolution Spectroscopic Survey (LRS), and constructed the photospheric and chromospheric basal flux calibrated stellar magnetic activity indictor.
And the Kepler \citep{2010Sci...327..977B} and Transiting Exoplanet Survey Satellite (TESS) \citep{2014SPIE.9143E..20R} also provided sufficient information of rotation period \citep{2014ApJS..211...24M, 2021ApJS..255...17S, 2023A&A...678A..24R, 2023ApJS..268....4F}.
Therefore, there are sufficient cross data to investigate the activity-rotation relationship. In this work, we focus on the relationship between the rotation and the chromospheric activity of solar-like stars.

In Section \ref{sec:data}, we introduced the selection criteria and composition of LAMOST solar-like stars. Section \ref{sec:calculation} illustrates how we estimate the chromospheric activity indictor based on the LAMOST LRS. We describe the chromospheric database and exhibit the rotation–activity relationship in Section \ref{sec:result}. Finally, we summarize the results and provide the conclusion in Section \ref{sec:summary}.

\section{Data Collection of Solar-like Stars} \label{sec:data}
\begin{figure*} 
    \begin{center}
    {\includegraphics[width=0.6\textwidth]{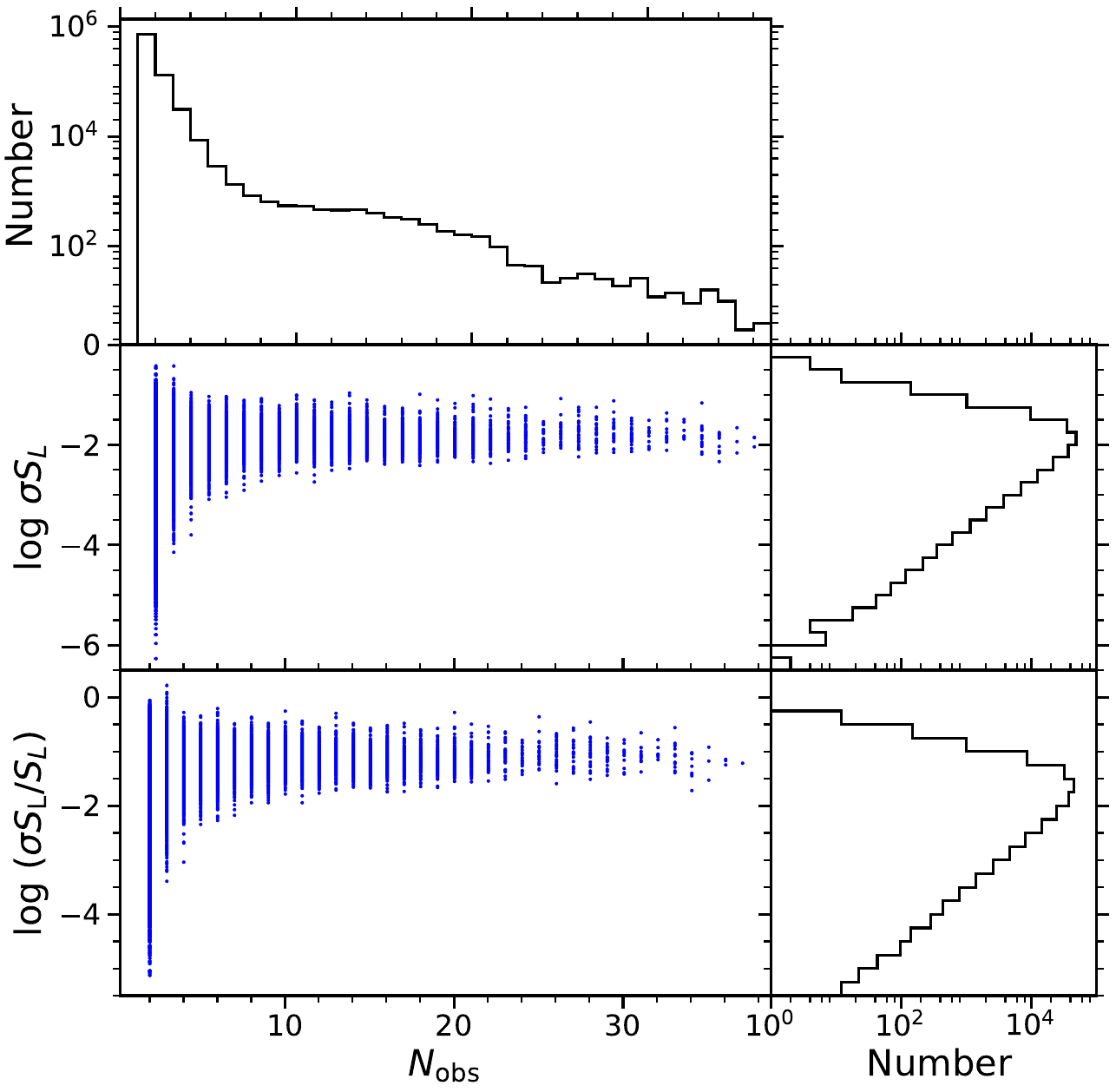}}   
    \end{center}

    \caption{Distribution of $\log\,\sigma S_L$ and $\log\,(\sigma S_L/S_L)$ versus the observation number. The top panel presents the histogram of observation counts, while the right panels display the corresponding distributions of $\log\,\sigma S_L$ (top right) and $\log\,(\sigma S_L/S_L)$ (bottom right).
    }
    \label{fig:S_L_std-Nobs}
\end{figure*}

The LAMOST Low-Resolution Spectroscopic Survey (LRS) provides spectroscopic data with a resolving power ($R=\lambda/\Delta\lambda$) of about 1800 and wavelength coverage of 3700--9100\,{\AA} \citep{2012RAA....12..723Z}. 
This work utilizes the LAMOST LRS spectra from Data Release 11 (DR11) v2.0 recorded in {\tt\string LAMOST LRS Stellar Parameter Catalog of A, F, G, and K Stars}\footnote{\url{http://www.lamost.org/dr11/v2.0/}}, which includes 7,898,024 spectra with determined stellar parameters observed between October 2011 and June 2023.
Fundamental stellar parameters such as effective temperature ($T_{\rm eff}$), surface gravity ($\log\,g$), metallicity ([Fe/H]), heliocentric radial velocity ($V_r$) along with their corresponding uncertainties, are provided by the LAMOST Stellar Parameter Pipeline (LASP) \citep{2015RAA....15.1095L}.
The selection of high-S/N spectra for solar-like stars was based on the same criteria used by \citet{2022ApJS..263...12Z}, with thresholds of ${\rm S/N}_g \ge 50.00$ and ${\rm S/N}_r \ge 71.43$.
The LAMOST spectra sample of solar-like stars is selected based on solar $T_{\rm eff}$ (5777\,K), $\log\,g$ (4.44\,dex) and [Fe/H] (0.0\,dex) with $4800 \le T_{\rm eff} \le 6300\,{\rm K}$, $-1.0<\text{[Fe/H]}<1.0\,{\rm dex}$ and $\log\,g \ge 5.98-0.00035 \times T_{\rm eff}$ the same as \citet{2024A&A...688A..23Z}.
When spectral flux data exhibit anomalies, we remove them by applying a simple criterion: any data point with a normalized flux greater than 1.2 or less than 1.0 is discarded as \citet{2024A&A...691A.218S}.
we eventually peak out 1,224,279 spectra (916,230 stars), in which 181,004 stars have been observed more than once. The distribution of the standard deviation of $S_L$ (see Section \ref{sec:calculation}) with the observation number can be seen in Figure \ref{fig:S_L_std-Nobs}. The distributions indicate that the standard deviation of $S_L$ is centered at approximately 0.018, and the relative standard deviation ($\sigma S_L/S_L$) is centered at approximately 0.056.

We acquired rotation periods from the Kepler and TESS missions by cross-matching our stellar sample ($1''$ tolerance) with the catalogs of \citet{2021ApJS..255...17S}, \citet{2023A&A...678A..24R} and \citet{2023ApJS..268....4F}.
we obtained 11,108 solar-like stars with chromospheric activity parameters and rotation period employed in Section \ref{sec:result}.

\section{Calculation of chromospheric activity index}\label{sec:calculation}

The emission in the cores of Ca\,\textsc{ii} H and K lines is measured by the $S$ index
defined as
\begin{equation} \label{eq:S_tri}
    S_{\rm tri} =  \frac{\widetilde{H}_{\rm tri} + \widetilde{K}_{\rm tri}}{\widetilde{R}+\widetilde{V}},
\end{equation}
where $\widetilde{H}_\mathrm{tri}$, $\widetilde{K}_\mathrm{tri}$, $\widetilde{R}$, and $\widetilde{V}$ are the mean flux measures in the corresponding H, K, R and V band centered in vacuum wavelengths 3969.59, 3934.78, 4002.20 3902.17 \AA, respectively \citep{2022ApJS..263...12Z, 2024A&A...688A..23Z}. The H and K bands are two 1.09 \AA\, full width at half maximum triangular band in the line core of Ca\,\textsc{ii} H and K lines, while the R and V bands are two 20 \AA\, rectangular band on both sides of the Ca\,\textsc{ii} H and K lines.
\citet{2016NatCo...711058K} employ $S_L$ to scale the $S$ index of LAMOST by
\begin{equation} \label{eq:S_L}
    S_L = \alpha_L \cdot \frac{8 \times 1.09\,\text{\AA}}{20\,\text{\AA}} \cdot S_{\rm tri},
\end{equation}
where the value of $\alpha_L$ is empirically determined as 1.8.
In our previous work, we calibrated the relationship between $S_L$ and $S_{\rm MWO}$ with an exponential function \citep{2022ApJS..263...12Z}. However, when $S_L$ is large, a deviation arises due to the limited coverage of the fitting data.
Using a more extensive set of LAMOST observations and cross-matching with the catalogs of \citet{1991ApJS...76..383D} and \citet{2018A&A...616A.108B} (see Appendix \ref{sec:S_calibration}), we have updated the empirical relation between $S_L$ and $S_{\rm MWO}$, which can be expressed as
\begin{equation} \label{eq:S_L_vs_S_MWO}
    S_{\rm MWO} = 2.747S_L-0.285,
\end{equation}

The $S_{\rm MWO}$ can be expressed by stellar surface flux as Equation \ref{eq:SMWO_vs_F}, where the dimensionless factor $\alpha$=2.4 \citep{1978PASP...90..267V, 1991ApJS...76..383D}.
\begin{equation} \label{eq:SMWO_vs_F}
    S_{\rm MWO} = 8\alpha \cdot \frac{\mathcal{F}_{\rm HK}}{\mathcal{F}_{\rm RV}}.
\end{equation}
The stellar surface flux normalized by bolometric flux is empirically evaluated by
\begin{equation} \label{eq:RHK_vs_FHK}
    R_{\rm HK} = \frac{\mathcal{F}_{\rm HK}}{\sigma T_{\rm eff}^4},
\end{equation}
where $\sigma = 5.67  \times 10^{-5}\,{\rm erg}\,{\rm cm}^{-2}\,{\rm s}^{-1}\,{\rm K}^{-4}$ is the Stefan-Boltzmann constant. Based on the Equations \ref{eq:SMWO_vs_F} an \ref{eq:RHK_vs_FHK}, the $R_{\rm HK}$ can be derived from $S_{\rm MWO}$ by
\begin{equation} \label{eq:RHK_vs_F_RV}
    R_{\rm HK} = \frac{ S_{\rm MWO} \cdot \mathcal{F}_{\rm RV}}{8\alpha} \cdot \frac{1}{\sigma T_{\rm eff}^4}.
\end{equation}
, where $\mathcal{F}_{\rm RV}$ can be estimated from empirical or synthetic spectral library \citep{ 1982A&A...107...31M,
1984A&A...130..353R,2015MNRAS.452.2745S,2017A&A...600A..13A,2018A&A...619A..73L,2023A&A...671A.162M,2024A&A...688A..23Z}. We estimate $\mathcal{F}_{\rm RV}$ based on the PHOENIX model \citep{2013A&A...553A...6H}. Details can be found in \citet{2024A&A...688A..23Z}.
The photospheric contribution can also be removed by
\begin{equation} \label{eq:Rp_HK}
    R'_{\rm HK} =\frac{\mathcal{F}_{\rm HK}-\mathcal{F}_{\rm phot}}{\sigma T_{\rm eff}^4} =R_{\rm HK} - R_{\rm phot},
\end{equation}
 where the $R_{\rm phot}$ can also be estimated from empirical or synthetic spectral library \citep{1984ApJ...276..254H, 1984ApJ...279..763N,2015MNRAS.452.2745S,2017A&A...600A..13A,2018A&A...619A..73L,2023A&A...671A.162M,2024A&A...688A..23Z}. Here we follow the method in \citet{1984ApJ...279..763N} to calculate the value of $R_{\rm phot}$ and calculate the value of $R'_{\rm HK}$ as Equations \ref{eq:Rphot-classic} and \ref{eq:teff_vs_bv}. 
 \begin{equation}\label{eq:Rphot-classic}
\log\,R_{\rm phot} = -4.898+1.918(B-V)^2-2.893(B-V)^3, \quad B-V>0.44
\end{equation}
\begin{equation} \label{eq:teff_vs_bv}
\log\,T_{\rm eff} = 3.908 - 0.234(B-V),\quad 0.4 < B-V < 1.4.
\end{equation}
 In Figure \ref{fig:RpHK_vs_RpHK_paper}, we compare the the values of classical chromospheric activity indicator $R'_{\rm HK}$ derived in this work with those reported in previous studies ($1''$ tolerance).
Furthermore, the chromospheric basal flux is thought to be a universal feature \citep{1987A&A...172..111S, 2014MNRAS.445..270P, 2013A&A...549A.117M}. The chromospheric basal flux removed $R$ index can be calculated by
\begin{equation}
\label{eq:R+_HK}
    R^+_{\rm HK} = \frac{\mathcal{F}_{\rm HK}-\mathcal{F}_{\rm phot}-\mathcal{F}_{\rm basal}}{\sigma T_{\rm eff}^4} = R_{\rm HK} - R_{\rm phot} - R_{\rm basal}.
\end{equation}

\begin{figure*} 
    \resizebox{\hsize}{!}{\includegraphics{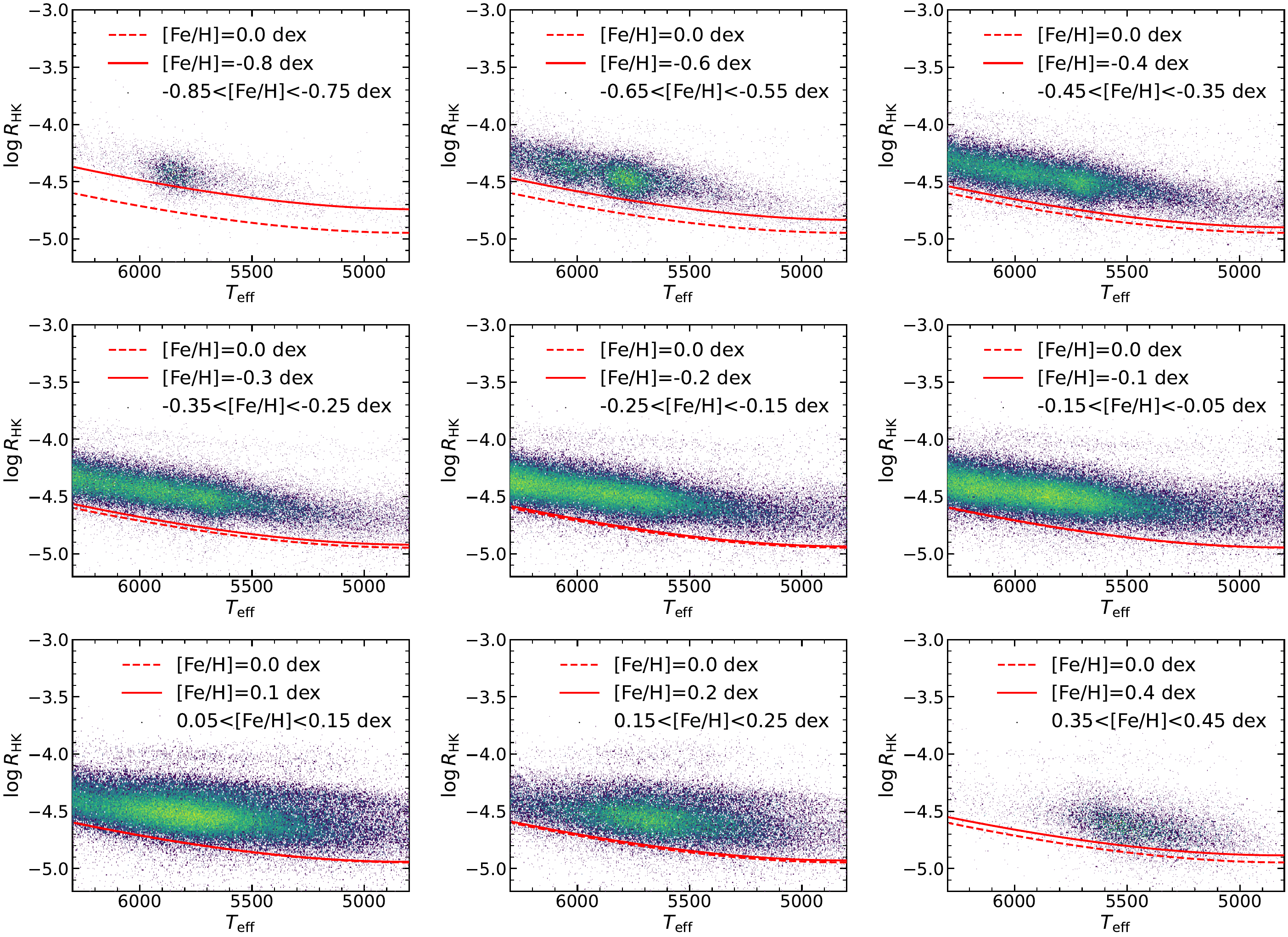}}
    \caption{Distribution of $\log R'_{\rm HK}$ versus $T_{\rm eff}$ for different [Fe/H] ranges. The color bar indicates the number density. The red solid line represents the results of Equation \ref{eq:R_phot_basal_vs_teff-feh} using the median [Fe/H] value for each metallicity range, while the red dotted line corresponds to the solar metallicity case ([Fe/H] = 0) for comparison.
    }
    \label{fig:R_HK-teff}
\end{figure*}

The basal chromospheric flux and the photospheric flux collectively define the lower boundary of $R_{\rm HK}$, which we denote as $R_{\rm phot,basal}$ ($R_{\rm phot,basal}=R_{\rm phot} + R_{\rm basal}$).
To model this boundary, we adopt a binary quadratic function in terms of $T_{\rm eff}$ and [Fe/H].
The stellar sample is partitioned into 30,000 bins in the $T_{\rm eff}$–[Fe/H] plane, with bin widths of 10 K in $T_{\rm eff}$ and 0.01 dex in [Fe/H].  Before fitting, the discrete data points were cleaned of outliers using the interquartile range (IQR) method.
$R_{\rm phot,basal}$ is obtained by fitting the minimum $R'_{\rm HK}$ in each bin with the median $T_{\rm eff}$ and [Fe/H] of that bin, expressed as
\begin{equation}\label{eq:R_phot_basal_vs_teff-feh}
\begin{aligned}
    \log R_{\rm phot,basal} 
    &\ =  332.27 -183.21 X +0.96 Y \\
    &\ -0.26 X \cdot Y + 24.88 X^{2} + 0.34 Y^{2},
\end{aligned}
\end{equation}
where $X$ and $Y$ represent $\log\,T_{\rm eff}$ and [Fe/H], respectively. Effective temperature dominates the variation in $R_{\rm phot,basal}$, while metallicity exhibits a weak influence, especially within the range of -0.2 to 0.2 dex, as demonstrated in Figure \ref{fig:R_HK-teff}.
Finally, we estimate $R^+_{\rm HK,L}$ by
\begin{equation}
\label{eq:R+_HK_LAMOST}
    R^+_{\rm HK,L} = R_{\rm HK} - R_{\rm phot,basal}.
\end{equation}

\section{Results and Discussion}\label{sec:result}

\subsection{The chromospheric database}

The LAMOST $S$ index $S_L$, the scaled $S$ index $S_{\rm MWO}$, the bolometric calibrated index $R_{\rm HK}$, the photospheric calibrated index $R'_{\rm HK}$, and the photospheric and chromospheric basal flux calibrated index $R^+_{\rm HK,L}$ introduced in Section \ref{sec:calculation} are provided in our database\footnote{\url{https://doi.org/10.5281/zenodo.18213069}}, stored in a CSV format file. 
In this work, We also provide the observation number and spectral information list for each solar-like stars observed more than once.
The maximum, minium, standard deviation values of the chromospheric activity parameters $S_L$, $S_{\rm MWO}$, $R_{\rm HK}$, $R'_{\rm HK}$ and $R^+_{\rm HK,L}$ for the solar-like stars observed more than once are also available. The detailed information of the database is shown in Table \ref{tab:catalog-columns}.

\startlongtable
\begin{deluxetable}{llllll}
	\tablecaption{Columns in the catalog of the database.\label{tab:catalog-columns}}
	\tablehead{
		& \colhead{Column} && \colhead{Unit} && \colhead{Description} 
	}
	\startdata
    & {\tt\string uid} &&  && Unique source identifier provided by LAMOST\\
    & {\tt\string ra} && degree && Right ascension (RA)\\
	& {\tt\string dec} && degree && Declination (DEC)\\
    & {\tt\string gp\_id} &&  && Source identifier in Pan-STARRS, Gaia or LAMOST\\
    & {\tt\string obs\_number} &&  && observation number\\
	& {\tt\string obsid\_list} &&  &&  List of unique spectrum ID\\
    & {\tt\string mjd\_list} &&  && List of modified Julian day\\
	& {\tt\string teff\_median} && K && Effective temperature ($T_\mathrm{eff}$)\\
	& {\tt\string teff\_std} && K && Standard deviation of $T_\mathrm{eff}$\\
	& {\tt\string logg\_median} && dex && Surface gravity ($\log\,g$) \\
	& {\tt\string logg\_std} && dex && Standard deviation of $\log\,g$ \\
	& {\tt\string feh\_median} && dex && Metallicity ([Fe/H]) \\
	& {\tt\string feh\_std} && dex && Standard deviation of [Fe/H] \\
	& {\tt\string rv\_median} && km/s && Radial velocity ($V_r$)\\
	& {\tt\string rv\_std} && km/s && Standard deviation of $V_r$\\
    & {\tt\string S\_L\_min} && && Minimum value of $S_L$ \\
	& {\tt\string S\_L\_max} && && Maximum value of $S_L$ \\
	& {\tt\string S\_L\_median} &&  && Median value of $S_L$ \\
	& {\tt\string S\_L\_std} && && Standard deviation of $S_L$\\
	& {\tt\string S\_MWO\_min} && && Minimum value of $S_{\rm MWO}$ \\
	& {\tt\string S\_MWO\_max} && && Maximum value of $S_{\rm MWO}$ \\    
	& {\tt\string S\_MWO\_median} && && Median value of $S_{\rm MWO}$ \\
	& {\tt\string S\_MWO\_std} && && Standard deviation of $S_{\rm MWO}$\\
	& {\tt\string R\_HK\_min} && && Minimum value of $R_{\rm HK}$ \\
	& {\tt\string R\_HK\_max} && && Maximum value of $R_{\rm HK}$ \\
	& {\tt\string R\_HK\_median} && && Median value of $R_{\rm HK}$ \\
	& {\tt\string R\_HK\_std} && && Standard deviation of $R_{\rm HK}$\\
	& {\tt\string Rp\_HK\_min} && && Minimum value of $R'_{\rm HK}$ \\
	& {\tt\string Rp\_HK\_max} && && Maximum value of $R'_{\rm HK}$ \\
	& {\tt\string Rp\_HK\_median} && && Median value of $R'_{\rm HK}$ \\
	& {\tt\string Rp\_HK\_std} && && Standard deviation of $R'_{\rm HK}$\\
	& {\tt\string R+\_HK\_min} && && Minimum value of $R^+_{\rm HK,L}$ \\
	& {\tt\string R+\_HK\_max} && && Maximum value of $R^+_{\rm HK,L}$ \\
	& {\tt\string R+\_HK\_median} && && Median value of $R^+_{\rm HK,L}$ \\
	& {\tt\string R+\_HK\_std} && && Standard deviation of $R^+_{\rm HK,L}$\\
	\enddata
\end{deluxetable}

\subsection{The relationship between chromospheric activity and the rotation period}\label{sec:CA_Prot}

We cross-match our samples ($1''$ tolerance) with the catalogs of \citet{2021ApJS..255...17S}, \citet{2023A&A...678A..24R} and \citet{2023ApJS..268....4F} to collect the rotation period.
we obtain 11,108 solar-like stars with chromospheric activity parameters and rotation period, which can be found in Table \ref{tab:catalog-prot}. The rotation period of TESS in \citet{2023ApJS..268....4F} is only taken from their Table 3, where the periods are strict and best characterized with the autocorrelation function (ACF).

\begin{figure*} 
    \resizebox{\hsize}{!}{\includegraphics{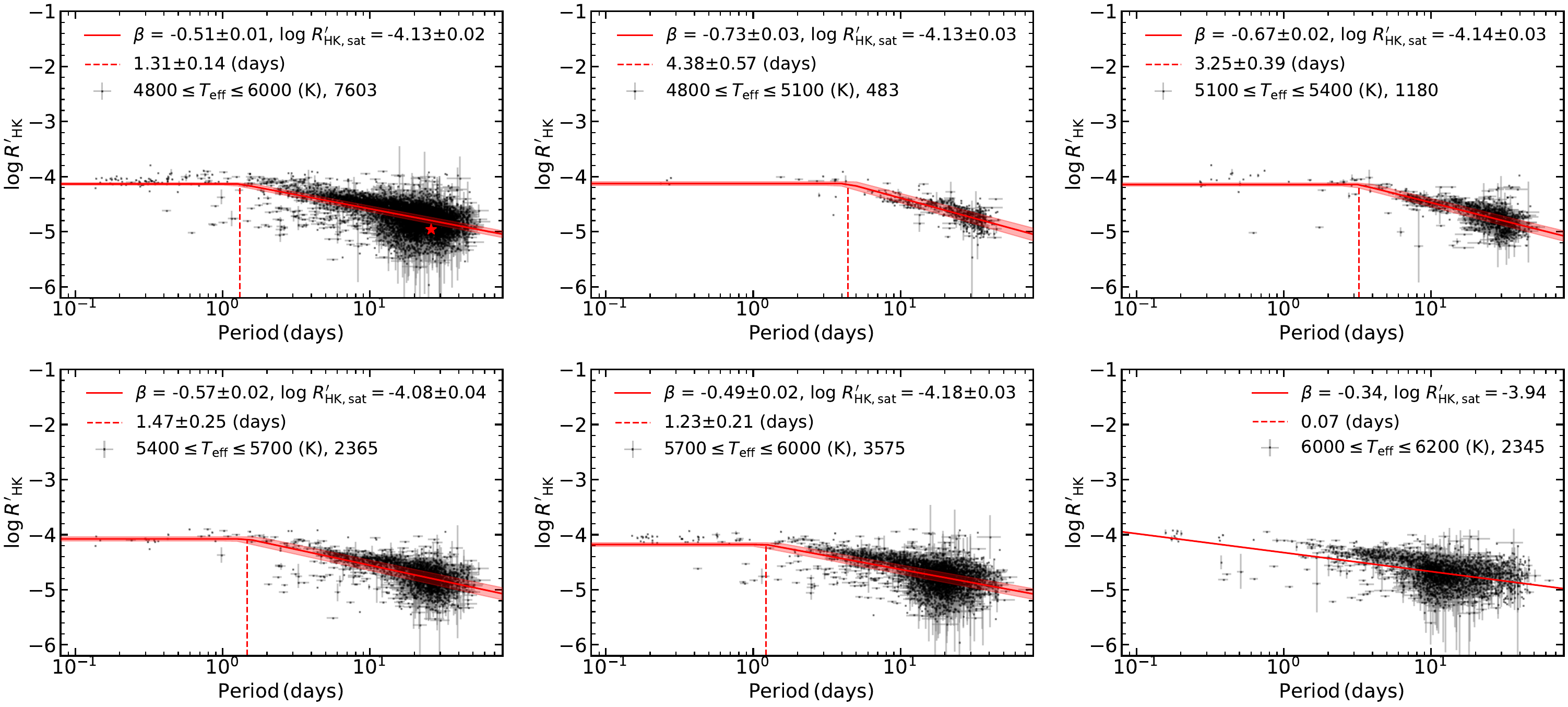}}
    \caption{Distribution of $\log R'_{\rm HK}$ versus rotation period for different $T_{\rm eff}$ ranges. We mark the position of the Sun with a red $\star$ symbol for comparison. The red dotted line indicates the position of $P_{\rm rot,sat}$ and the red solid line is the fitted segmented function line. The error bars of $\log R'_{\rm HK}$ represent the standard deviations for stars with repeated measurements.
    }
    \label{fig:Rp_HK_Prot_teff}
\end{figure*}

\begin{figure*} 
    \resizebox{\hsize}{!}{\includegraphics{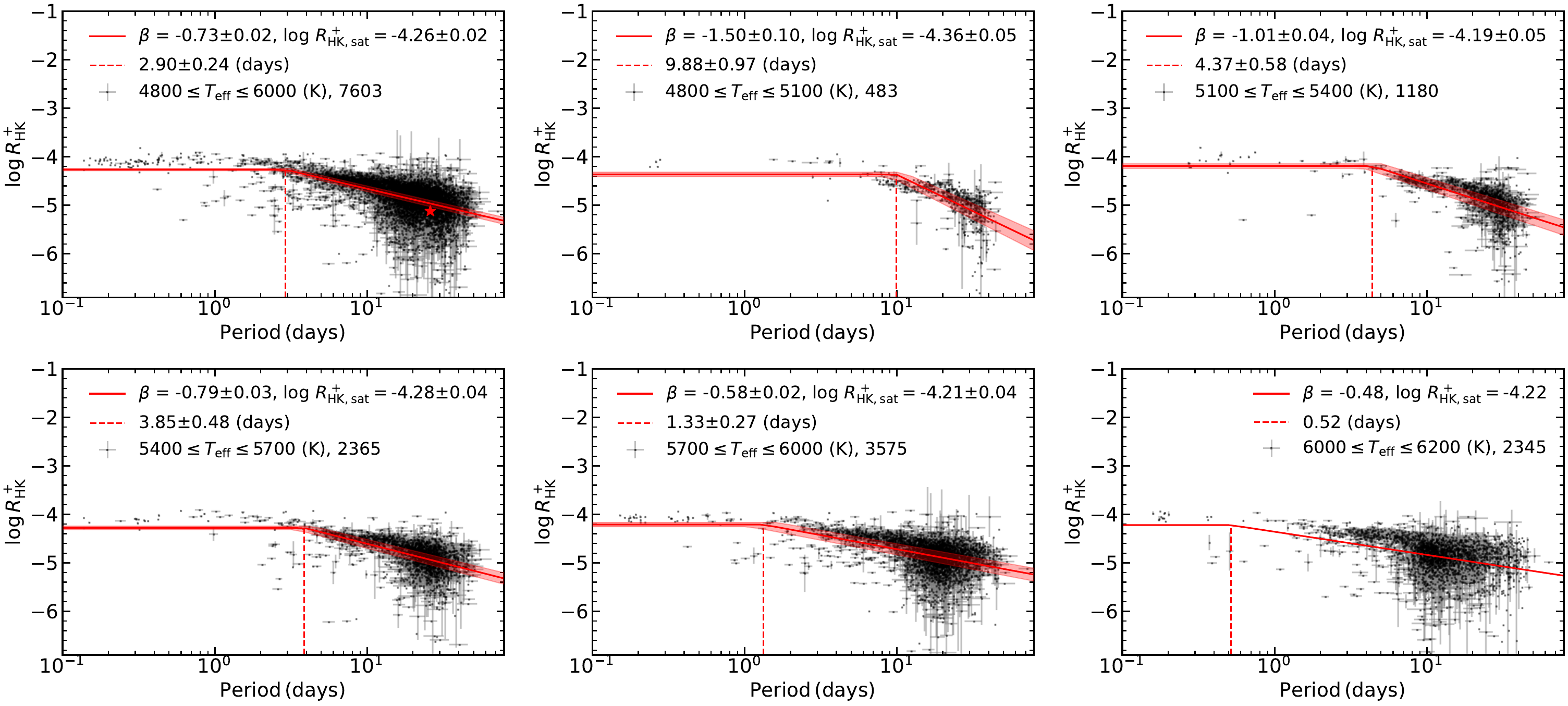}}
    \caption{Distribution of $\log R^+_{\rm HK,L}$ versus rotation period for different $T_{\rm eff}$ ranges, sames as Figure \ref{fig:Rp_HK_Prot_teff}.
    }
    \label{fig:Rc_HK_Prot_teff}
\end{figure*}

The value of $T_{\rm eff}$ can also be found in \citet{2021ApJS..255...17S} and \citet{2023ApJS..268....4F}.
After applying a 100 K cut to the difference between $T_{\rm eff,median}$ in our database and $T_{\rm eff}$ in these studies, we recomputed the relevant parameters in Figures \ref{fig:Rp_HK_Prot_teff} and \ref{fig:Rc_HK_Prot_teff}. The resulting values are consistent with the original parameters within their respective uncertainties.
Cross-matching our sample of 11,108 stars with \citet{2025A&A...696A.243G} (Gaia) and \citet{2025ApJS..276...57G} (TESS), of which 8,556 are not classified as binary systems in the reference catalogs.
The corresponding values in Figures \ref{fig:Rp_HK_Prot_teff} and \ref{fig:Rc_HK_Prot_teff} derived from the 8,556 stars are also consistent with the parameters discussed later, within their respective uncertainties. Therefore, in the following discussion, we examine the activity-rotation relation using the 11,108 samples and the associated stellar atmospheric parameters from LAMOST.

We employ the piecewise function
\begin{equation}\label{eq:R-Prot-4para}
\log R = 
\begin{cases}
\log R_{\rm sat},  & P_{\rm rot} < P_{\rm rot,sat} \\
\beta \cdot \log P_{\rm rot} + C, & P_{\rm rot} \geq P_{\rm rot,sat}
\end{cases}
\end{equation}
to describe the relationship of both $R'_{\rm HK}$ and $R^+_{\rm HK,L}$ to the rotation period.
By requiring the function to be continuous, the constant $C$ can be solved as $\log R_{\rm sat} - \beta \cdot \log P_{\rm rot,sat}$.
Therefore, Equation \ref{eq:R-Prot-4para} can be simplified as a function of three undetermined parameters ($\log R_{\rm sat}$, $\beta$ and $P_{\rm rot,sat}$) as
\begin{equation}\label{eq:R-Prot-3para}
\log R = 
\begin{cases}
\log R_{\rm sat},  & P_{\rm rot} < P_{\rm rot,sat} \\
\beta \cdot \log P_{\rm rot} + \log R_{\rm sat} - \beta \cdot \log P_{\rm rot,sat}, & P_{\rm rot} \geq P_{\rm rot,sat}
\end{cases}.
\end{equation}

The sufficient sample size enables us to examine the activity-rotation correlation in various stellar parameter intervals.
Firstly, we investigate the relationship between $\log R'_{\rm HK}$ and $\log P_{\rm rot}$.
We show the activity-rotation distribution and the fitted activity-rotation distribution relationship for stars in different $T_{\rm eff}$ ranges, as can be seen in Figure \ref{fig:Rp_HK_Prot_teff}.
We use the nonlinear least square method through the python module {\tt\string curve\_fit} of {\tt\string scipy.optimize} \citep{2020NatMe..17..261V} to estimate the values of $\log R_{\rm sat}$, $P_{\rm rot,sat}$ , $\beta$ and their corresponding uncertainties in Equation \ref{eq:R-Prot-3para}.

Saturation of chromospheric activity is commonly detected, but the evidence becomes less reliable for stars with $T_{\rm eff}$ greater than 6000 K. 
The corresponding parameters $\log R'_{\rm HK,sat}$, $P_{\rm rot,sat}$ and $\beta$ are poorly constrained, as indicated by uncertainties that are much greater than the parameter values.
The spin-down process ceases for stars with $T_{\rm eff}\gtrsim$6200 K due to the disappearance of the surface convection zone \citep{1967ApJ...150..551K, 2020ApJ...888...43C}.
Given the gentler slope (smaller $\beta$) and the absence of saturation data at $T_{\rm eff}>$6000 K, the fit naturally converges on a linear function.
The absence of saturation data at $T_{\rm eff}>$6000 K has two possible explanations: insufficient data coverage, or the absence of the saturation phenomenon in F-type stars as \citep{2011ApJ...743...48W}.

The slope value $\beta$ and the saturated value of $P_{\rm rot}$ generally decrease with increasing $T_{\rm eff}$. 
As the stellar effective temperature increases from 4950 to 5850 K, the slope $\beta$ and $P_{\rm rot,sat}$ vary correspondingly from -0.73 to -0.49 and from 4.38 to 1.23 days, respectively.
And the values of $\beta$ and $P_{\rm rot}$ for $4800 \leq T_{\rm eff} \leq 6000$ K are close to the case of $5400 \leq T_{\rm eff} \leq 5700$ K and $5700 \leq T_{\rm eff} \leq 6000$ K, around the solar effective temperature. The solar values of $\log R'_{\rm HK}$ and $P_{\rm rot}$ take as -4.960 and 26 days, respectively  \citep{2024A&A...688A..23Z}.

Similarly as in the case of $\log R'_{\rm HK}$, we estimate the relationship between $\log R^+_{\rm HK,L}$ and $\log P_{\rm rot}$.
When fitting data for solar-like stars across different effective temperature ranges, the saturation of chromospheric activity also becomes less reliable for $T_{\rm eff}>$ 6000 K.
The slope value $\beta$ and the saturation value of $P_{\rm rot}$ also decrease with increasing $T_{\rm eff}$, similar to the behavior observed for $\log R'_{\rm HK}$ (as shown in Figure \ref{fig:Rc_HK_Prot_teff}). 
The slope $\beta$ and $P_{\rm rot,sat}$ for $4800 \leq T_{\rm eff} \leq 6000$ K are close to the case of $5400 \leq T_{\rm eff} \leq 5700$ K and $5700 \leq T_{\rm eff} \leq 6000$ K. The solar values of $\log R^+_{\rm HK,L}$ take as -5.121 calculated based on Equation \ref{eq:R+_HK_LAMOST}. 
For stars in different $T_{\rm eff}$ ranges, the values of $P_{\rm rot,sat}$ and $\beta$ for the $\log R^+_{\rm HK,L}$ indicator are systematically larger than those for the $\log R'_{\rm HK}$ indicator (see Table \ref{tab:Prot-parameters}). 
The higher slope $\beta$ of the unsaturated activity-rotation relationship indicates a more rapid change in magnetic activity with rotation. 
Therefore, we conclude that the $\log R^+_{\rm HK,L}$ indicator is more sensitive to rotation variation than the $\log R'_{\rm HK}$ indicator.
For stars in the $T_{\rm eff}$ range of 4950 to 5850 K, the saturation rotation periods range from 9.88 to 1.33 days, while the unsaturated regime slopes vary from -1.50 to -0.58.

\begin{figure*} 
    \resizebox{\hsize}{!}{\includegraphics{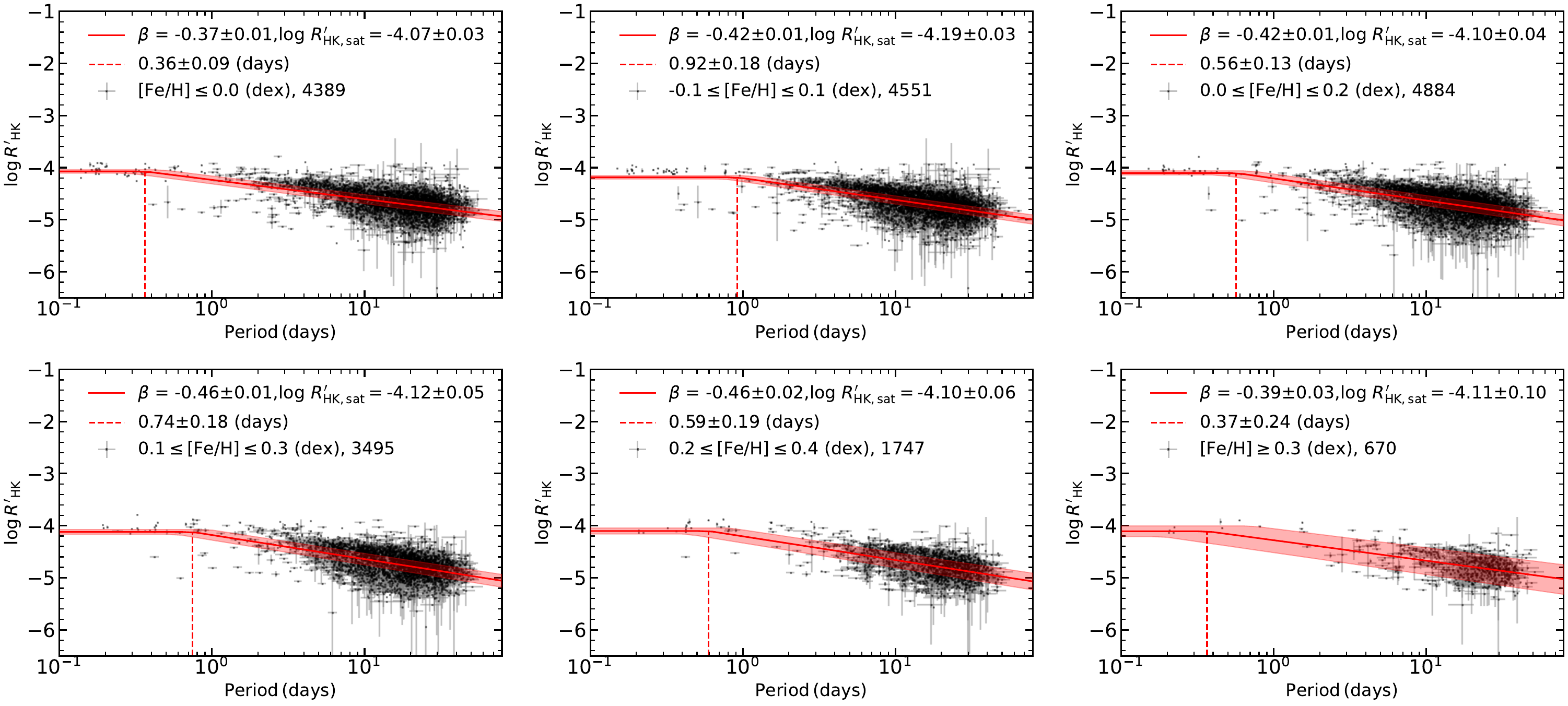}}
    \caption{Distribution of $\log R'_{\rm HK}$ versus rotation period for different [Fe/H] ranges, sames as Figure \ref{fig:Rp_HK_Prot_teff}. 
    }
    \label{fig:Rp_HK_Prot_feh}
\end{figure*}

\begin{figure*} 
    \resizebox{\hsize}{!}{\includegraphics{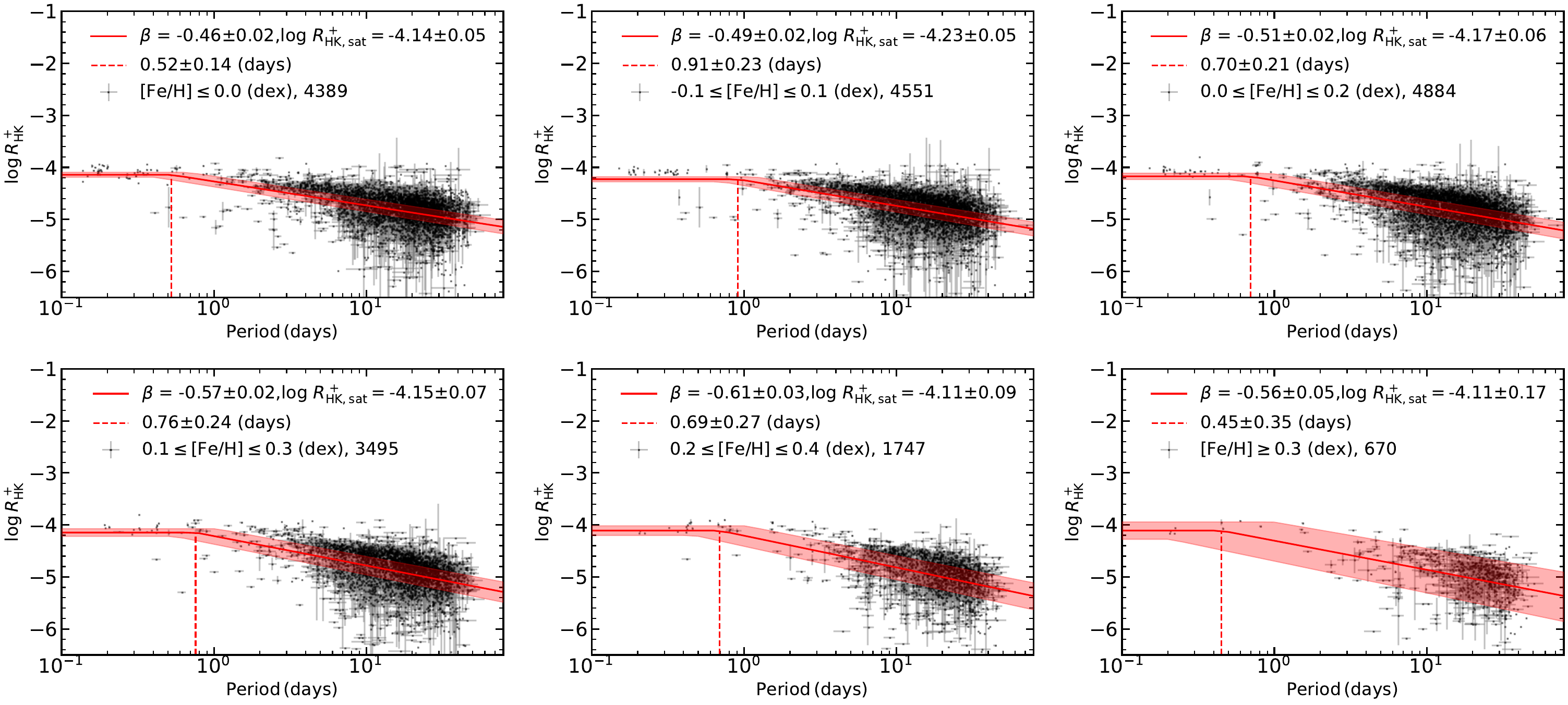}}
    \caption{Distribution of $\log R^+_{\rm HK}$ versus rotation period for different [Fe/H] ranges, sames as Figure \ref{fig:Rp_HK_Prot_teff}. 
    }
    \label{fig:Rc_HK_Prot_feh}
\end{figure*}

An increase in metallicity leads to higher stellar opacity. This increased opacity deepens the surface convection zone, which in turn raises the convective overturn timescale \citep{2020ApJ...888...43C}.
Based on the stars in $T_{\rm eff}$ range of 4800 to 6000 K, We also study the relationship between $\log R^+_{\rm HK,L}$ and $\log P_{\rm rot}$ for different [Fe/H] ranges.
The distribution of $\log R'_{\rm HK}$ vs. $\log P_{\rm rot}$ and $\log R^+_{\rm HK,L}$ vs. $\log P_{\rm rot}$ are shown in Figures \ref{fig:Rp_HK_Prot_feh} and \ref{fig:Rc_HK_Prot_feh}, respectively.
We also provided the fitting parameters $\beta$, $P_{\rm rot,sat}$ and $\log R_{\rm ,sat}$ fro different [Fe/H] ranges, which are summarized in Table \ref{tab:Prot-parameters}.
The maximum values of $\beta$ approximately tend appear for [Fe/H] in the range of 0.2 to 0.3 dex for both $\log R'_{\rm HK}$ and $\log R^+_{\rm HK,L}$.

\begin{deluxetable}{cllllll}
\tablecaption{The parameters of $\beta$, $P_{\rm rot,sat}$ and $\log R_{\rm sat}$ in Equation \ref{eq:R-Prot-3para} (activity-$P_{\rm rot}$) for different $T_{\rm eff}$ and [Fe/H] ranges. \label{tab:Prot-parameters}}
\tablehead{
\colhead{Range of} & \multicolumn{2}{c}{$\beta$}  & \multicolumn{2}{c}{$P_{\rm rot,sat}$  (days)} & \multicolumn{2}{c}{$\log R_{\rm sat}$}\\
\cline{2-7}
\colhead{$T_{\rm eff}$ (K)} & \colhead{$\log R'_{\rm HK}$} & \colhead{ $\log R^+_{\rm HK,L}$} & \colhead{$\log R'_{\rm HK}$} & \colhead{$\log R^+_{\rm HK,L}$} & \colhead{$\log R'_{\rm HK}$} & \colhead{$\log R^+_{\rm HK,L}$}
}
\startdata
4800,6000 & -0.51$\pm$0.01 & -0.73$\pm$0.02 & 1.31$\pm$0.14 & 2.90$\pm$0.24 & -4.13$\pm$0.02 & -4.26$\pm$0.02 \\
4800,5100 & -0.73$\pm$0.03 & -1.50$\pm$0.10 & 4.38$\pm$0.57 & 9.88$\pm$0.97 & -4.13$\pm$0.03 & -4.36$\pm$0.05 \\
5100,5400 & -0.67$\pm$0.02 & -1.01$\pm$0.04 & 3.25$\pm$0.39 & 4.37$\pm$0.58 & -4.14$\pm$0.03 & -4.19$\pm$0.05 \\
5400,5700 & -0.57$\pm$0.02 & -0.79$\pm$0.03 & 1.47$\pm$0.25 & 3.85$\pm$0.48 & -4.08$\pm$0.04 & -4.28$\pm$0.04 \\
5700,6000 & -0.49$\pm$0.02 & -0.58$\pm$0.02 & 1.23$\pm$0.21 & 1.33$\pm$0.27 & -4.18$\pm$0.03 & -4.21$\pm$0.04 \\
\hline
[Fe/H] (dex) & & & & & & \\
\hline
$\leq$0.0 & -0.37$\pm$0.01 & -0.46$\pm$0.02 & 0.36$\pm$0.09 & 0.52$\pm$0.14 & -4.07$\pm$0.03 & -4.14$\pm$0.05 \\
-0.1,0.1 & -0.42$\pm$0.01 & -0.49$\pm$0.02 & 0.92$\pm$0.18 & 0.91$\pm$0.23 & -4.19$\pm$0.03 & -4.23$\pm$0.05 \\
0.0,0.2 & -0.42$\pm$0.01 & -0.51$\pm$0.02 & 0.56$\pm$0.13 & 0.70$\pm$0.21 & -4.10$\pm$0.04 & -4.17$\pm$0.06 \\
0.1,0.3 & -0.46$\pm$0.01 & -0.57$\pm$0.02 & 0.74$\pm$0.18 & 0.76$\pm$0.24 & -4.12$\pm$0.05 & -4.15$\pm$0.07 \\
0.2,0.4 & -0.46$\pm$0.02 & -0.61$\pm$0.03 & 0.59$\pm$0.19 & 0.69$\pm$0.27 & -4.10$\pm$0.06 & -4.11$\pm$0.09 \\
$\geq$0.3 & -0.39$\pm$0.03 & -0.56$\pm$0.05 & 0.37$\pm$0.24 & 0.45$\pm$0.35 & -4.11$\pm$0.10 & -4.11$\pm$0.17 \\
\enddata
\end{deluxetable}

\subsection{The relationship between chromospheric activity and the Rossby number}\label{sec:CA-Ro}

The Rossby number ${\rm Ro} = P_{\rm rot}/\tau_c$ is generally used to study the relationship between chromospheric activity and the rotation period \citep{1984ApJ...279..763N}.
The $P_{\rm rot}$ are taken from \citet{2021ApJS..255...17S}, \citet{2023A&A...678A..24R} and \citet{2023ApJS..268....4F}. 
The convective turnover time $\tau_c$ can be determined empirically \citep{1984ApJ...279..763N, 2003A&A...397..147P, 2011ApJ...743...48W, 2018MNRAS.479.2351W} or theoretically \citep{1998A&A...334..953V, 2017ApJ...838..161S}.
\citet{2003A&A...397..147P} demonstrated that their empirical formulation agreed well with \citep{1984ApJ...279..763N}, especially in the color range $0.5<(B-V)<1.0$. This color index range is consistent with the effective temperature range (4800-6300 K) of our stellar sample.
Therefore, we employ the formula in \citet{1984ApJ...279..763N} to estimate the Rossby number, expressed as
\begin{equation}\label{eq:tau_bv}
\tau_c = 
\begin{cases}
1.362-0.166x+0.025x^2-5.323x^3,  & x>0 \\
1.362-014x, & x<0
\end{cases},
\end{equation}
where $x=1-(B-V)$. The values of $B-V$ are calculated based on Equation \ref{eq:teff_vs_bv}.
Using Equation \ref{eq:R-Prot-3para} introduced in Section \ref{sec:CA_Prot}, we describe the relationship of both $R'_{\rm HK}$ and $R^+_{\rm HK,L}$ with Ro by
\begin{equation}\label{eq:R-Ro-3para}
\log R = 
\begin{cases}
\log R_{\rm sat},  & \log {\rm Ro} < \log {\rm Ro_{\rm sat}} \\
\beta \cdot \log {\rm Ro} + \log R_{\rm sat} - \beta \cdot \log {\rm Ro_{\rm sat}}, & \log {\rm Ro} \geq \log {\rm Ro_{\rm sat}}
\end{cases}.
\end{equation}

\begin{figure*} 
    \resizebox{\hsize}{!}{\includegraphics{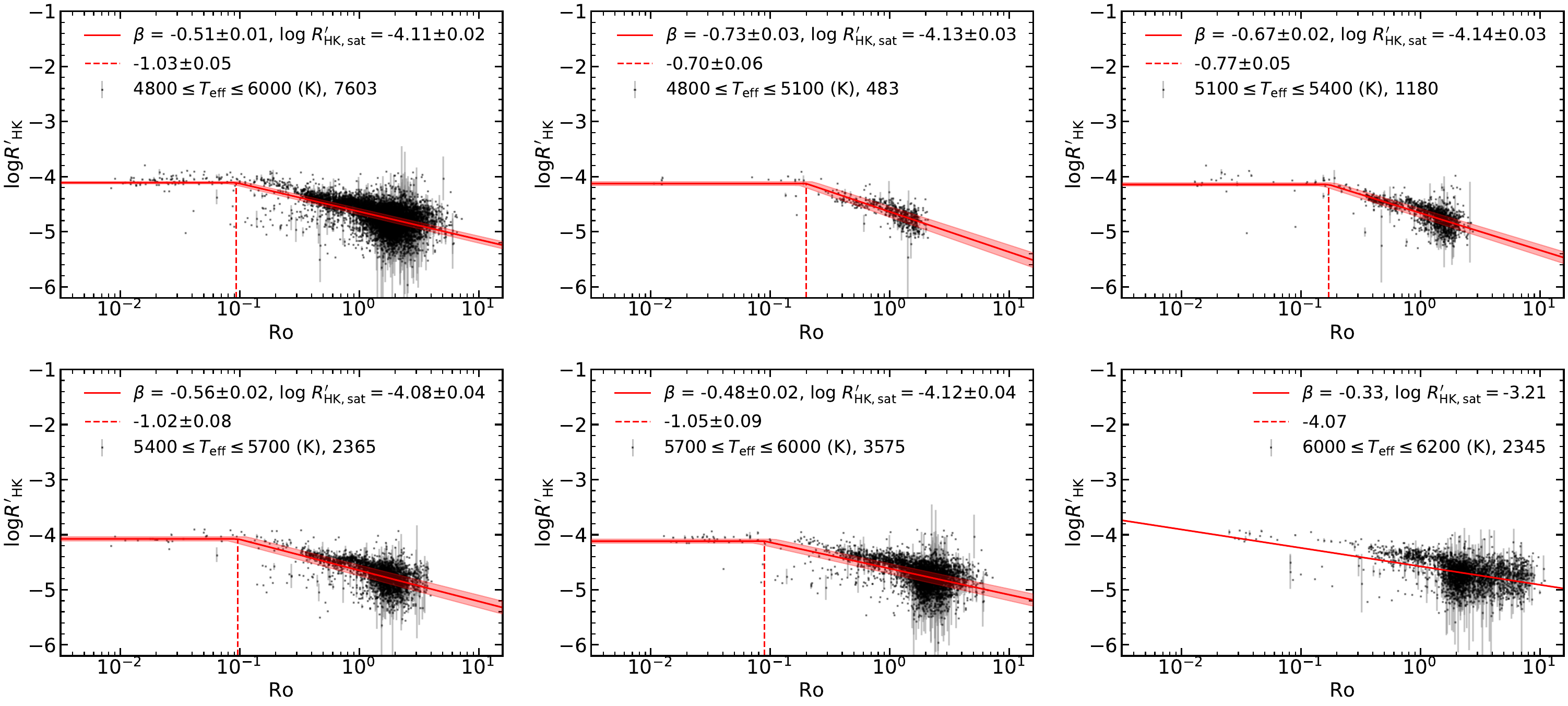}}
    \caption{Distribution of $\log R'_{\rm HK}$ versus Ro for different $T_{\rm eff}$ ranges, same as Figure \ref{fig:Rp_HK_Prot_teff}.
    }
    \label{fig:Rp_HK_Ro_teff}
\end{figure*}

\begin{figure*} 
    \resizebox{\hsize}{!}{\includegraphics{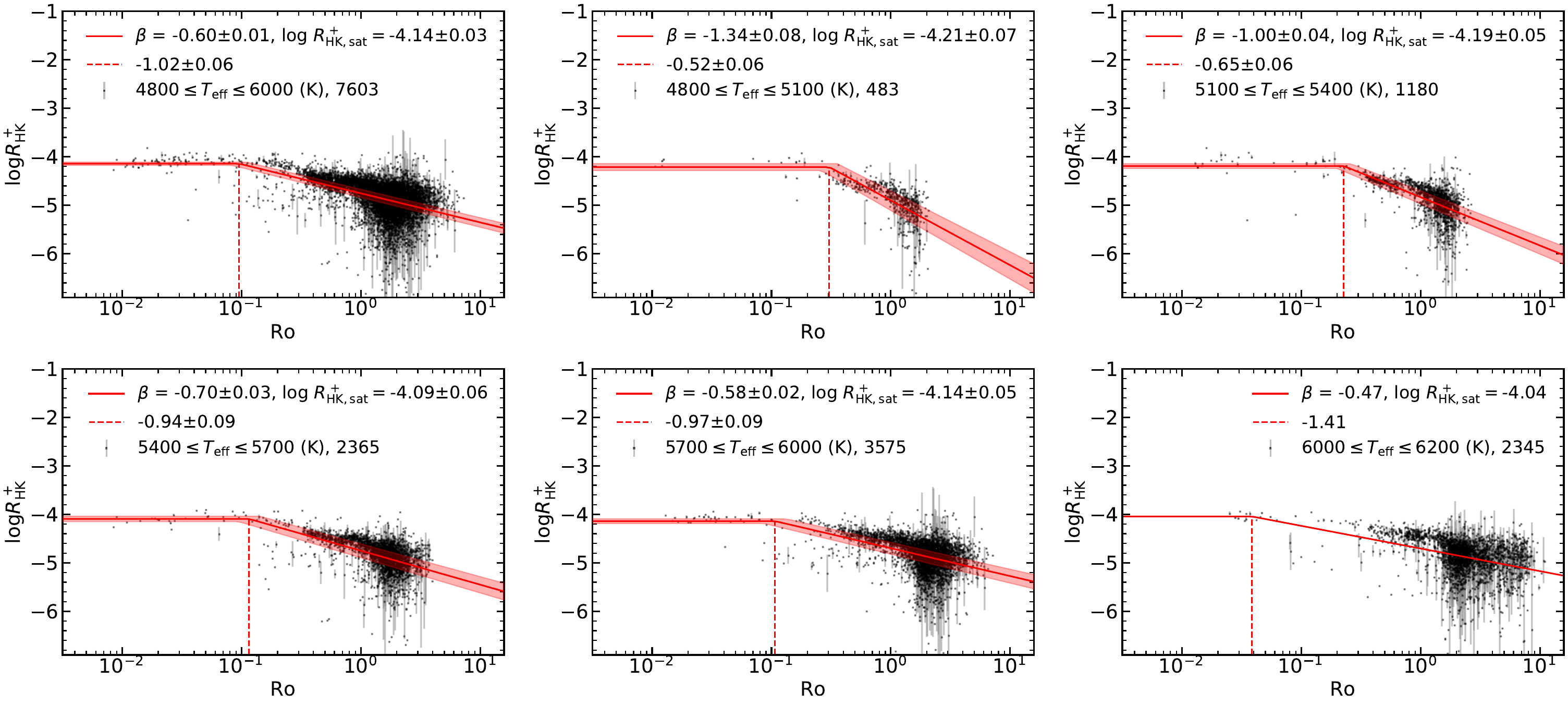}}
    \caption{Distribution of $\log R^+_{\rm HK,L}$ versus Ro for different $T_{\rm eff}$ ranges, same as Figure \ref{fig:Rp_HK_Prot_teff}.
    }
    \label{fig:Rc_HK_Ro_teff}
\end{figure*}

Due to the small value of Ro, we enlarged Ro by a factor of 100 during fitting. The slope value and the saturated value of Ro also decrease with increasing $T_{\rm eff}$ for both the case of $R'_{\rm HK}$ and $R^+_{\rm HK,L}$ (as shown in Figures \ref{fig:Rp_HK_Ro_teff} and \ref{fig:Rc_HK_Ro_teff}). The values of slope $\beta$ of the unsaturated activity-rotation relation and $\log R_{\rm ,sat}$ derived using the Ro across different $T_{\rm eff}$ ranges, is generally consistent with the slope derived using $P_{\rm rot}$.

For stars in different $T_{\rm eff}$ ranges, the values of both the slope and the saturated Ro for the $\log R^+_{\rm HK,L}$ indicator are consistently greater than those for the $\log R'_{\rm HK}$ indicator, as can be seen in Table \ref{tab:Ro-parameters}. 
For $T_{\rm eff}$ in the range of 4950 to 5850 K, the saturated Ro is in the range of 0.200-0.031 and 0.302-0.107 for $R'_{\rm HK}$ and $R^+_{\rm HK,L}$, respectively.
The value of $\rm Ro_{sat}$ is generally consistent with other studies \citep{2003A&A...397..147P, 2011ApJ...743...48W, 2025ApJS..281....5Z}.
\citet{2025ApJS..281....5Z} concluded that the saturation of H$\alpha$ line occurs at $\rm Ro<0.130$ ($\rm Ro/Ro_{\odot}<0.07$, $\rm Ro_{\odot}=1.85$) for K-type stars.
Different methods used to estimate $\tau_c$ may result in discrepancies in the values corresponding to the saturation regime \citep{2025ApJS..281....5Z}.

For stars with $T_{\rm eff}$ in the range of 4950 to 5850 K, the values of slop $\beta$ in the unsaturated regime vary from -0.73 to -0.48 and from -1.34 to -0.58 for $R'_{\rm HK}$ and $R^+_{\rm HK,L}$, respectively.
The variation of $\beta$ with Ro is similar to that with $P_{\rm rot}$ in Section \ref{sec:CA_Prot} for both $R'_{\rm HK}$ and $R^+_{\rm HK,L}$, as shown in Table \ref{tab:Prot-parameters} and Table \ref{tab:Ro-parameters}.
In summary, the slope $\beta$ of the unsaturated activity-rotation relation, derived using the Ro across different $T_{\rm eff}$ ranges, is generally consistent with the slope derived using $P_{\rm rot}$.
For both the case of $\log R'_{\rm HK}$-Ro and $\log R^+_{\rm HK,L}$-Ro, the values of $\rm  Ro_{sat}$ and the slope $\beta$ for the broad $T_{\rm eff}$ range of 4800-6000 K closely match those derived for the ranges of 5400-5700 K and 5700-6000 K, which are near the solar $T_{\rm eff}$.

\begin{figure*} 
    \resizebox{\hsize}{!}{\includegraphics{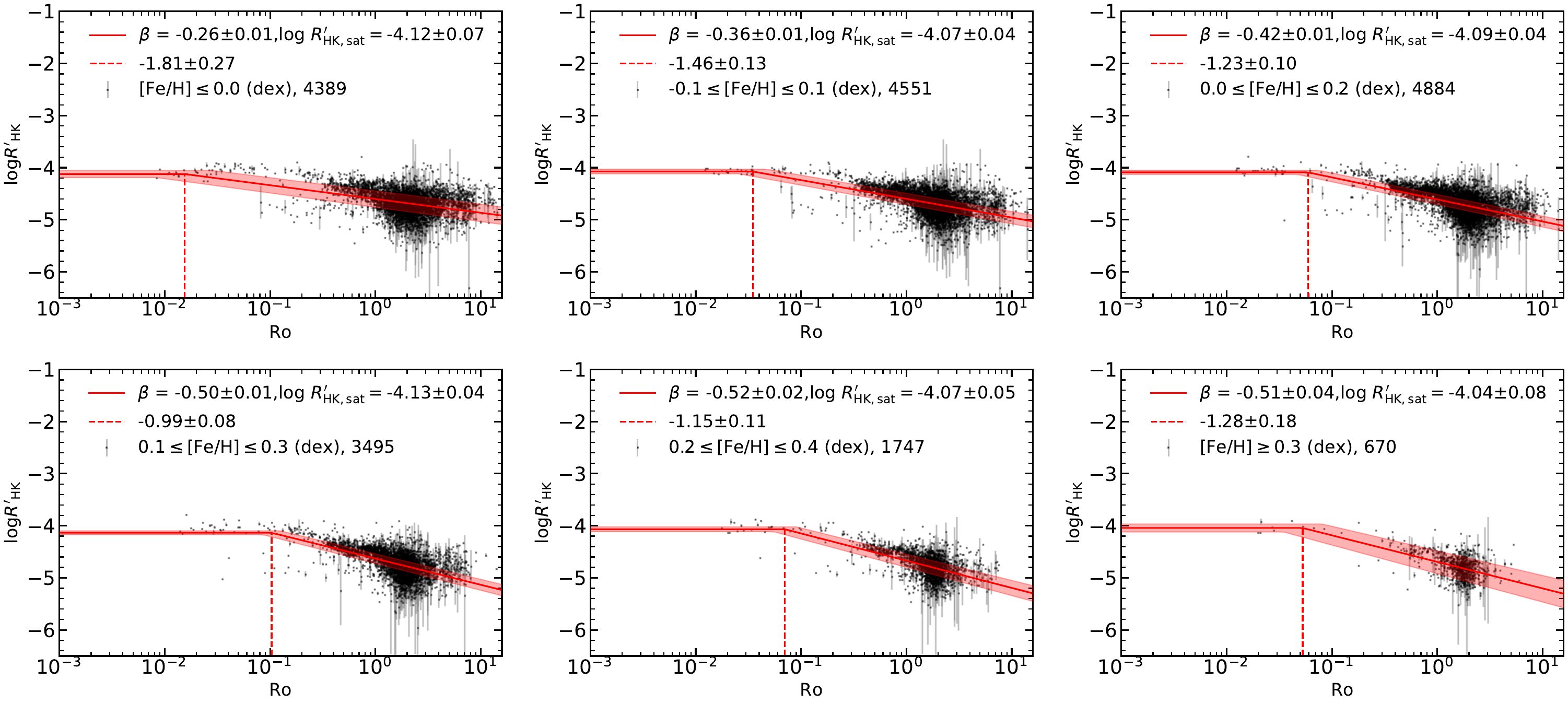}}
    \caption{Distribution of $\log R^+_{\rm HK}$ versus rotation period for different [Fe/H] ranges, sames as Figure \ref{fig:Rp_HK_Prot_teff}. 
    }
    \label{fig:Rp_HK_Ro_feh}
\end{figure*}

\begin{figure*} 
    \resizebox{\hsize}{!}{\includegraphics{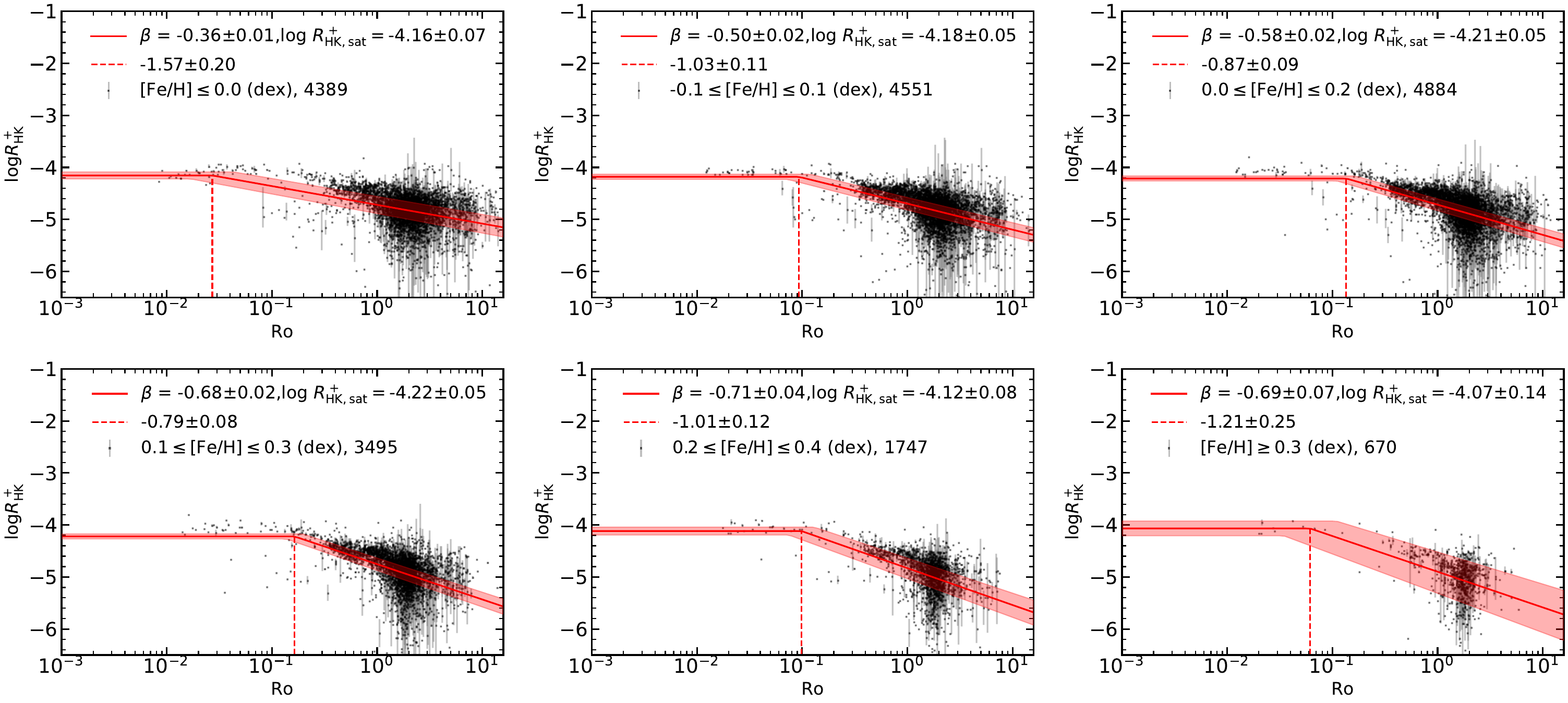}}
    \caption{Distribution of $\log R^+_{\rm HK}$ versus rotation period for different [Fe/H] ranges, sames as Figure \ref{fig:Rp_HK_Prot_teff}. 
    }
    \label{fig:Rc_HK_Ro_feh}
\end{figure*}

Based on the stars in $T_{\rm eff}$ range of 4800 to 6000 K, We also study the relationship between $\log R^+_{\rm HK,L}$ and $\log {\rm Ro}$ for different [Fe/H] ranges (as shown in Figures \ref{fig:Rp_HK_Ro_feh} and \ref{fig:Rc_HK_Ro_feh}).
The distribution of $\log R'_{\rm HK}$ vs. $\log P_{\rm rot}$ and $\log R^+_{\rm HK,L}$ vs. $\log P_{\rm rot}$ are shown in Figures \ref{fig:Rp_HK_Prot_feh} and \ref{fig:Rc_HK_Prot_feh}, respectively.
The parameters $\beta$, $\rm Ro_{sat}$ and $\log R_{\rm ,sat}$ for different [Fe/H] can be seen in Table \ref{tab:Ro-parameters}. 
The value of $\beta$ becomes relatively stable when [Fe/H]$>$0.1 dex for both $\log R'_{\rm HK}$ and $\log R^+_{\rm HK,L}$. 
The $\rm Ro_{sat}$ are located in the range of 0.015 to 0.102 and the range of 0.027 to 0.162 for $\log R'_{\rm HK}$ and $\log R^+_{\rm HK,L}$, respectively.
The maximum value of $\rm Ro_{sat}$ appear for [Fe/H]$\approx$0.2 dex.

\begin{deluxetable}{cllllll}
\tablecaption{The parameters of $\beta$, $P_{\rm rot,sat}$ and $\log R_{\rm sat}$ in Equation \ref{eq:R-Ro-3para} (activity-Ro) for different $T_{\rm eff}$ and [Fe/H] ranges. \label{tab:Ro-parameters}}
\tablehead{
\colhead{Range of} & \multicolumn{2}{c}{$\beta$}  & \multicolumn{2}{c}{log Ro} & \multicolumn{2}{c}{$\log R_{\rm sat}$}\\
\cline{2-7}
\colhead{$T_{\rm eff}$ (K)} & \colhead{$\log R'_{\rm HK}$} & \colhead{ $\log R^+_{\rm HK,L}$} & \colhead{$\log R'_{\rm HK}$} & \colhead{$\log R^+_{\rm HK,L}$} & \colhead{$\log R'_{\rm HK}$} & \colhead{$\log R^+_{\rm HK,L}$}
}
\startdata
4800,6000 & -0.51$\pm$0.01 & -0.60$\pm$0.01 & -1.03$\pm$0.05 & -1.02$\pm$0.06 & -4.11$\pm$0.02 & -4.14$\pm$0.03 \\
4800,5100 & -0.73$\pm$0.03 & -1.34$\pm$0.08 & -0.70$\pm$0.06 & -0.52$\pm$0.06 & -4.13$\pm$0.03 & -4.21$\pm$0.07 \\
5100,5400 & -0.67$\pm$0.02 & -1.00$\pm$0.04 & 0.77$\pm$0.05 & -0.65$\pm$0.06 & -4.14$\pm$0.03 & -4.19$\pm$0.05 \\
5400,5700 & -0.56$\pm$0.02 & -0.70$\pm$0.03 & -1.02$\pm$0.08 & -0.94$\pm$0.09 & -4.08$\pm$0.04 & -4.09$\pm$0.06 \\
5700,6000 & -0.48$\pm$0.02 & -0.58$\pm$0.02 & -1.50$\pm$0.09 & -0.97$\pm$0.09 & -4.12$\pm$0.04 & -4.14$\pm$0.05 \\
\hline
[Fe/H] (dex) & & & & & & \\
\hline
$\leq$0.0 & -0.26$\pm$0.01 & -0.36$\pm$0.01 & -1.81$\pm$0.27 & -1.57$\pm$0.20 & -4.12$\pm$0.07 & -4.16$\pm$0.07 \\
-0.1,0.1 & -0.36$\pm$0.01 & -0.50$\pm$0.02 & -1.46$\pm$0.13 & -1.03$\pm$0.11 & -4.07$\pm$0.04 & -4.18$\pm$0.05 \\
0.0,0.2 & -0.42$\pm$0.01 & -0.58$\pm$0.02 & -1.23$\pm$0.10 & -0.87$\pm$0.09 & -4.09$\pm$0.04 & -4.21$\pm$0.05 \\
0.1,0.3 & -0.50$\pm$0.01 & -0.68$\pm$0.05 & -0.99$\pm$0.08 & -0.79$\pm$0.08 & -4.13$\pm$0.04 & -4.22$\pm$0.05 \\
0.2,0.4 & -0.52$\pm$0.02 & -0.71$\pm$0.04 & -1.15$\pm$0.11 & -1.01$\pm$0.12 & -4.07$\pm$0.05 & -4.12$\pm$0.08 \\
$\geq$0.3 & -0.51$\pm$0.04 & -0.69$\pm$0.07 & -1.28$\pm$0.18 & -1.21$\pm$0.25 & -4.04$\pm$0.08 & -4.07$\pm$0.14 \\
\enddata
\end{deluxetable}

\subsection{Activity-rotation correlation analysis for stars with multiple observations}

\begin{figure*} 
    \resizebox{\hsize}{!}{\includegraphics{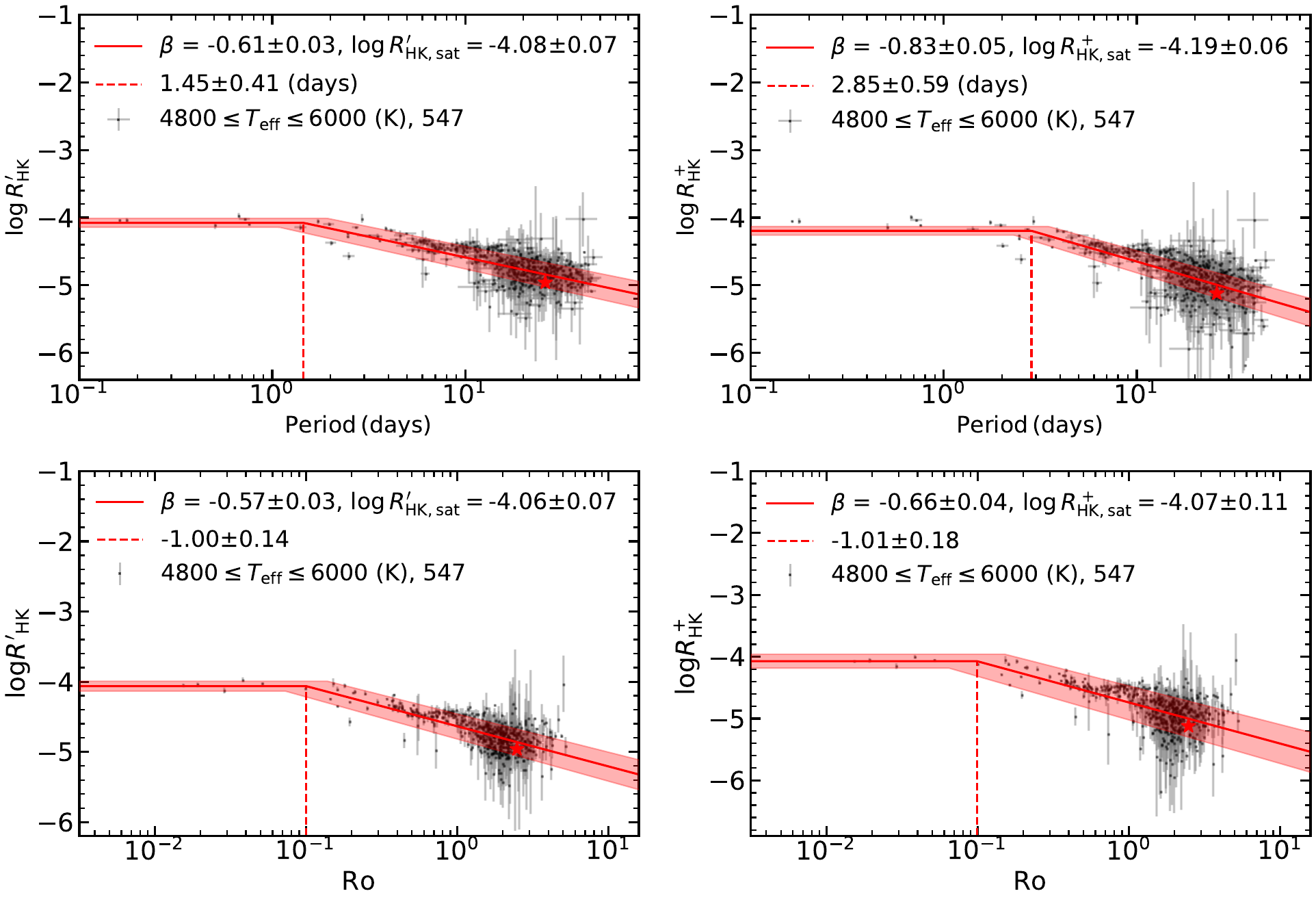}}
    \caption{Distribution of $\log R'_{\rm HK}$ versus $P_{\rm rot}$ (upper left), $\log R^+_{\rm HK,L}$ versus $P_{\rm rot}$ (upper right), $\log R'_{\rm HK}$ versus Ro (bottom left) and $\log R^+_{\rm HK,L}$ versus Ro (bottom right) for stars observed more than twice. The red dotted line indicates the position of $P_{\rm rot,sat}$ and the red solid line is the fitted segmented function line. And the light red interval represents the error interval of the fitting curve.
    }
    \label{fig:Rp_HK(Rc_HK)_Prot(Ro)}
\end{figure*}

In Sections \ref{sec:CA_Prot} and \ref{sec:CA-Ro}, we introduced the relationship between Ca\,\textsc{ii} H and K lines emission with rotation period and Rossby number.
We investigate the activity-rotation relationship using stars with three or more observations to minimize the effects of measurement scatter.
Due to the lack of stars that have been observed multiple times by LAMOST and have rotation period parameters in other studies, we can currently only set the limit to three or more observations.
We exhibit the activity-rotation relation for stars ($4800 \leq T_{\rm eff} \leq 6000$ K) observed more than twice in Figure \ref{fig:Rp_HK(Rc_HK)_Prot(Ro)}.
The samples are almost all fall within the fitting interval.
the relationship between $R'_{\rm HK}$ and $P_{\rm rot}$ can be described as
\begin{equation}\label{eq:Rp_HK-prot}
\log R'_{\rm HK} = 
\begin{cases}
-4.08\pm0.07,  & P_{\rm rot} < 1.45\pm0.41 {\rm \, days} \\
(-0.61\pm0.03)\log P_{\rm rot} -3.98, & P_{\rm rot} \geq 1.45\pm0.41 {\rm \, days}
\end{cases},
\end{equation}
and the relationship between $R^+_{\rm HK,L}$ and $P_{\rm rot}$ is
\begin{equation}\label{eq:Rc_HK-prot}
\log R^+_{\rm HK,L} = 
\begin{cases}
-4.19\pm0.06,  & P_{\rm rot} < 2.85\pm0.59 {\rm \, days} \\
(-0.83\pm0.05)\log P_{\rm rot} -3.81, & P_{\rm rot} \geq 2.85\pm0.59 {\rm \, days}
\end{cases}.
\end{equation}
The saturation of chromospheric activity occurs at $P_{\rm rot}$=1.45 days and $P_{\rm rot}$=2.85 days, for the cases of $R'_{\rm HK}$ and $R^+_{\rm HK,L}$, respectively. 
The values of $P_{\rm rot,sat}$ and $R_{\rm sat}$ in Equations \ref{eq:Rp_HK-prot} and \ref{eq:Rc_HK-prot} are consistent with those in the first line of Table \ref{tab:Prot-parameters} within the range of their uncertainties.
The slop $\beta$ in the unsaturated region is $-0.61\pm0.03$ ($R'_{\rm HK}$) and $-0.83\pm0.05$ ($R^+_{\rm HK,L}$), while $-0.51\pm0.01$ ($R'_{\rm HK}$) and $-0.73\pm0.02$ ($R^+_{\rm HK,L}$) in Section \ref{sec:CA_Prot}.
The value of $\Delta \beta$ is 0.1, indicating no significant variation.

Similarly, the relationship of $R'_{\rm HK}$-Ro and $R^+_{\rm HK,L}$-Ro are expressed by Equations \ref{eq:Rp_HK-Ro} and \ref{eq:Rc_HK-Ro}.
The saturation of chromospheric activity indicators occurs at Ro=0.100 and Ro=0.097, for the cases of $R'_{\rm HK}$ and $R^+_{\rm HK,L}$, respectively.
The values of $\beta$, $\rm Ro_{sat}$ and $\log R_{\rm sat}$ in Equations \ref{eq:Rp_HK-Ro} and \ref{eq:Rc_HK-Ro} are consistent with those in the first line of Table \ref{tab:Ro-parameters} within the range of their uncertainties.
This consistency suggests that stellar variability does not exert a significant influence on the fits presented in Figures \ref{fig:Rp_HK_Prot_teff}-\ref{fig:Rc_HK_Ro_feh}.
Several vertically scattered data points deviate from the fitted range, potentially due to factors such as stellar structure, evolutionary characteristics, long-term magnetic cycles, or uncertainties in the spectra.
The fitting interval primarily covers regions of high data density, suggesting minimal influence from vertical distribution.

\begin{equation}\label{eq:Rp_HK-Ro}
\log R'_{\rm HK} = 
\begin{cases}
-4.06\pm0.07,  & \log {\rm Ro} < -1.00\pm0.14 \\
(-0.57\pm0.03)\log {\rm Ro} -4.63, & \log {\rm Ro} \geq -1.00\pm0.14
\end{cases}
\end{equation}
\begin{equation}\label{eq:Rc_HK-Ro}
\log R^+_{\rm HK,L} = 
\begin{cases}
-4.07\pm0.11,  & \log {\rm Ro} < -1.01\pm0.18 \\
(-0.66\pm0.04)\log {\rm Ro} -4.74, & \log {\rm Ro} \geq -1.01\pm0.18
\end{cases}
\end{equation}

\section{Summary and Conclusion} \label{sec:summary}

In this work, we provide the chromospheric activity database for 916,230 solar-like stars observed by LAMOST DR11. The LAMOST $S$ index $S_L$, the scaled $S$ index $S_{\rm MWO}$, the bolometric calibrated index $R_{\rm HK}$, the photospheric calibrated index $R'_{\rm HK}$, and the photospheric and chromospheric basal flux calibrated index $R^+_{\rm HK,L}$ are available in our database. We also provide the values of minium, maximum and standard deviation of these activity indicators for 181,004 solar-like stars observed more than once by LAMOST.

In our chromospheric activity database, 11,108 solar-like stars with rotation period observed by Kepler and TESS were picked out to investigate the activity-rotation relationship. Our statistical results show that the values of the chromospheric activity index $R'_{\rm HK}$ and $R^+_{\rm HK,L}$ increase with rotation rate until it reaches a saturation level. The downward trend and saturation are more significant in the photospheric and chromospheric basal flux calibrated index $R^+_{\rm HK,L}$ than in the photospheric calibrated index $R'_{\rm HK}$. 
While $T_{\rm eff}$ increases from 4950 to 5850 K, the saturated $P_{\rm rot}$ varies correspondingly from 4.38 to 1.23 days and from 9.88 to 1.33 days for $R'_{\rm HK}$ and $R^+_{\rm HK,L}$, respectively.

The statistical results of the Rossby number Ro are similar to those derived from $P_{\rm rot}$. 
The slope $\beta$ of the unsaturated activity-rotation relation, derived using the Ro across different $T_{\rm eff}$ ranges, is generally consistent with the slope derived using the $P_{\rm rot}$.
While $T_{\rm eff}$ increases from 4950 to 5850 K, the saturated Ro varies  correspondingly from 0.200 to 0.032 and from 0.302 to 0.107 for $R'_{\rm HK}$ and $R^+_{\rm HK,L}$, respectively. 
The downward trend and saturation for the $R^+_{\rm HK,L}$ indicator are also more significant than those for the $R'_{\rm HK}$ indicator.

We estimate the relationship between chromospheric activity and rotation for stars observed more than twice and in the $T_{\rm eff}$ range of 4800-6000 K.
For $R'_{\rm HK}$, the saturation occurs at $P_{\rm rot}$=1.45 days (Ro = 0.100), and the slop of unsaturated region is -0.61 (-0.57 for Ro).
For $R^+_{\rm HK,L}$, the saturation occurs at $P_{\rm rot}$=2.85 days (Ro = 0.097), and the slop of unsaturated region is -0.83 (-0.66 for Ro).

The statistical results of the activity-rotation relationship are shown by the chromospheric activity indicators $R'_{\rm HK}$ and $R^+_{\rm HK,L}$ derived from LAMOST in this work.
Although our study contains a substantial amount of stellar chromospheric activity data with crossed rotation periods, rapidly rotating stars remain relatively rare among solar-like stars. 
Given a fixed rotation period, cooler stars with deeper convective envelopes exhibit stronger rotational modulation, which influence the periodic detection fraction \citep{2014ApJS..211...24M, 2021ApJS..255...17S}.
Expanding the sample to include a broader range of stellar types in future searches will allow for a clearer revelation of the activity-rotation relationship across a wider range of effective temperatures.
Although the LAMOST spectroscopic survey has been conducted for about 13 years (comparable to a solar cycle), it remains difficult to characterize the temporal variations of stellar activity due to the limited number of observations per star.
Future work will involve identifying suitable stellar samples and using multiple observations to investigate the temporal characteristics of stellar activity.

\appendix{}

\section{Calibration of Chromospheric Activity Indicators}\label{sec:S_calibration}

\begin{figure*} 
    \begin{center}
    \includegraphics[width=0.50\textwidth]{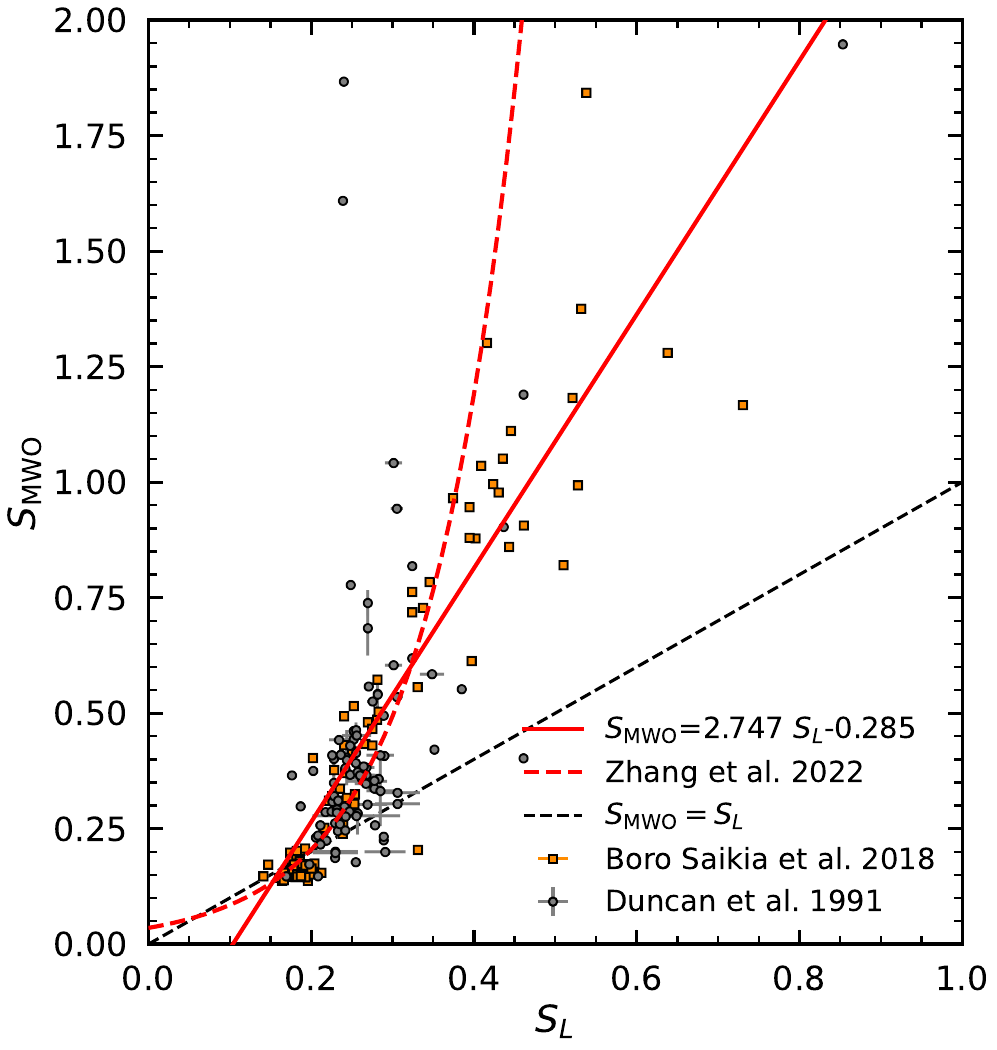}
    \end{center}
    \caption{The distribution of $S_L$ values derived from LAMOST spectra with the $S_{\rm MWO}$ values reported in previous studies. The black dotted line indicates $S_{\rm MWO}=S_L$, while the red solid line indicates the best-fit relationship between the two indices.
    }
    \label{fig:S_L_vs_S_MWO}
\end{figure*}

\begin{figure*} 
    \begin{center}
    \includegraphics[width=0.50\textwidth]{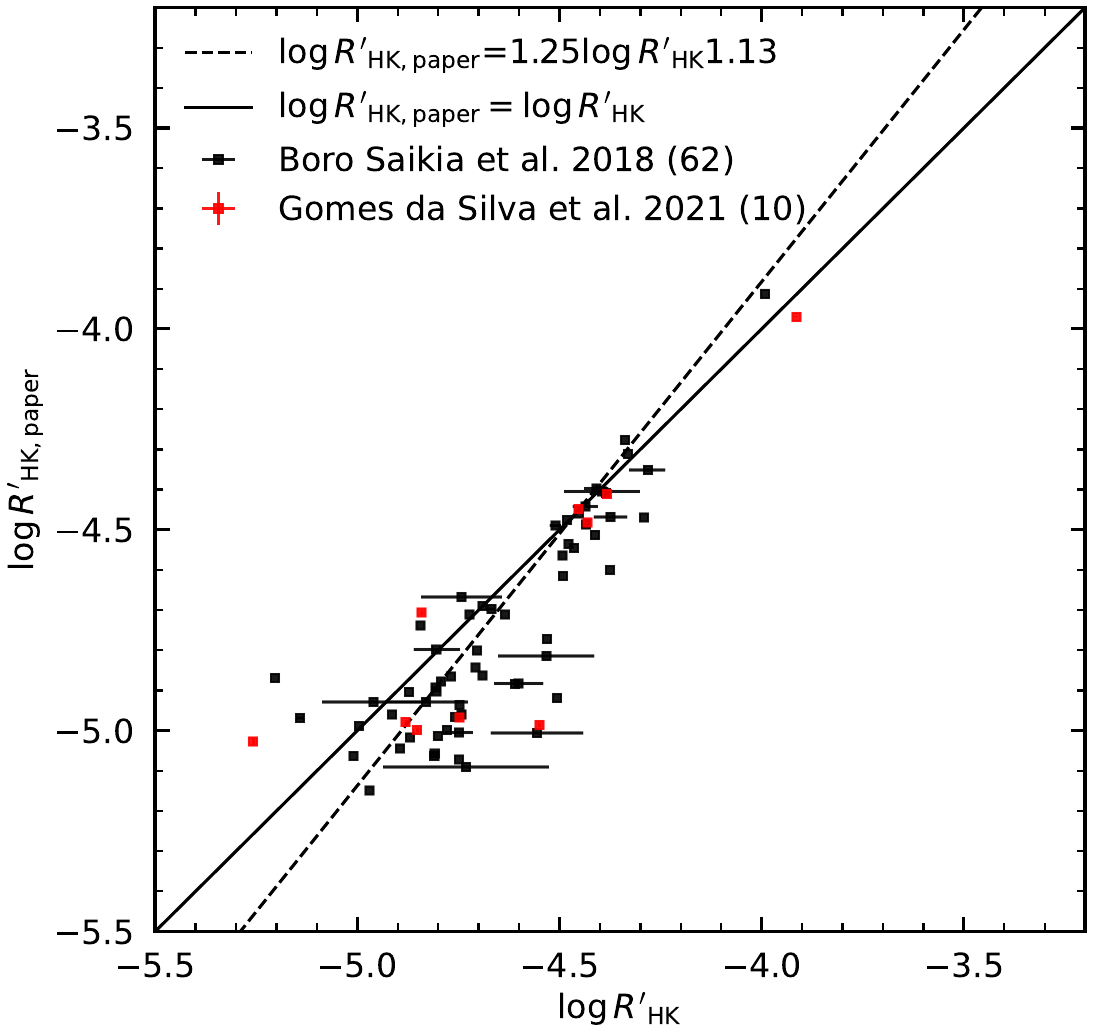}
    \end{center}
    \caption{The distribution of $\log R'_{\rm HK}$ values derived in this work with those reported in previous studies. The black solid line indicates $\log R'_{\rm HK}$= $\log R'_{\rm HK,paper}$, and the dotted line is their fitting line.
    }
    \label{fig:RpHK_vs_RpHK_paper}
\end{figure*}

We identified 194 common stars between the LAMOST LRS catalog and the $S_{\rm MWO}$ catalogs provided by \citet{1991ApJS...76..383D} (94 stars) and \citet{2018A&A...616A.108B} (100 stars), using a $1''$ cross-match tolerance.
A linear fit was performed to describe the relationship between $S_L$ and $S_{\rm MWO}$. This relation is applicable to our entire sample, with a maximum $S_L$ value of 0.92. 
The fitting results are presented in Figure \ref{fig:S_L_vs_S_MWO}, with the corresponding data available at \url{https://doi.org/10.5281/zenodo.18213069}. 
Furthermore, we compared the classical chromospheric activity indicator $R'_{\rm HK}$ derived in this work with values from \citet{2018A&A...616A.108B} and \citet{2021A&A...646A..77G} (using a $1''$ cross-match tolerance), as shown in Figure \ref{fig:RpHK_vs_RpHK_paper}.

\section{The data of chromospheric activity and rotation}
The data of rotation period used to discuss the activity-rotation relationship in Section \ref{sec:result} is cross-matched from \citet{2021ApJS..255...17S, 2023A&A...678A..24R, 2023ApJS..268....4F} within the tolerance of 1 arcsec in coordinates.
The chromospheric activity indicators and the rotation period for the cross-matched solar-like stars are available online\footnote{\url{https://doi.org/10.5281/zenodo.18213069}}, detailed descriptions can be found in Table \ref{tab:catalog-prot}. 

\startlongtable
\begin{deluxetable}{llllll}
	\tablecaption{Columns in the catalog of the database.\label{tab:catalog-prot}}
	\tablehead{
		& \colhead{Column} && \colhead{Unit} && \colhead{Description} 
	}
	\startdata
    & {\tt\string uid} &&  && Unique source identifier provided by LAMOST\\
    & {\tt\string ra} && degree && Right ascension (RA)\\
	& {\tt\string dec} && degree && Declination (DEC)\\
    & {\tt\string gp\_id} &&  && Source identifier in Pan-STARRS, Gaia or LAMOST\\
    & {\tt\string obs\_number} &&  && observation number\\
	& {\tt\string teff\_median} && K && Effective temperature ($T_\mathrm{eff}$)\\
	& {\tt\string teff\_std} && K && Standard deviation of $T_\mathrm{eff}$\\
	& {\tt\string logg\_median} && dex && Surface gravity ($\log\,g$) \\
	& {\tt\string logg\_std} && dex && Standard deviation of $\log\,g$ \\
	& {\tt\string feh\_median} && dex && Metallicity ([Fe/H]) \\
	& {\tt\string feh\_std} && dex && Standard deviation of [Fe/H] \\
	& {\tt\string Rp\_HK\_median} && && Median value of $R'_{\rm HK}$ \\
	& {\tt\string Rp\_HK\_std} && && Standard deviation of $R'_{\rm HK}$\\
	& {\tt\string R+\_HK\_median} && && Median value of $R^+_{\rm HK,L}$ \\
	& {\tt\string R+\_HK\_std} && && Standard deviation of $R^+_{\rm HK,L}$\\
 	& {\tt\string Prot} && days && rotation period $P_{\rm rot}$\\
	& {\tt\string Prot\_err} && days && Uncertainty of $P_{\rm rot}$\\
    & {\tt\string Prot\_source} && days && Source of $P_{\rm rot}$\\
	\enddata
	\tablecomments{The parameters of solar-like stars with cross-matched rotation period.}
\end{deluxetable}

\begin{acknowledgements}

This work is supported by the National Natural Science Foundation of China (12573054).
Weitao Zhang acknowledges the support of the Nanxun Scholars Program for Young Scholars of ZJWEU (RC2024021500).
Han He was supported by the National Natural Science Foundation of China (11973059). The authors were supported by the National Key R\&D Program of China (2019YFA0405000). 
Guoshoujing Telescope (the Large Sky Area Multi-Object Fiber Spectroscopic Telescope, LAMOST) is a National Major Scientific Project built by the Chinese Academy of Sciences. Funding for the project has been provided by the National Development and Reform Commission. LAMOST is operated and managed by the National Astronomical Observatories, Chinese Academy of Sciences.
\end{acknowledgements}

\facility{LAMOST}
\software{Astropy \citep{2013A&A...558A..33A, 2018AJ....156..123A},
          SciPy \citep{2020NatMe..17..261V},
          NumPy \citep{2007CSE.....9c..10O, 2011CSE....13b..22V,2020Natur.585..357H},
          Matplotlib \citep{2007CSE.....9...90H}
          }

\bibliographystyle{aa}
\bibliography{ref}

\end{CJK*}
\end{document}